%% file: main.tex
\documentclass[a4paper,12pt]{article}
\input{misc/preamble}

\begin{document}

%-------------------- TITLE PAGE --------------
\input{front/title_page}

%------------------ Table of Contents ---------
\clearpage
\tableofcontents
\thispagestyle{empty}

%---------------------- Main Body ----------------------
\clearpage
\setcounter{page}{1}
\setcounter{footnote}{0}

\input{sections/1_introduction}
\input{sections/2_conventions}
\input{sections/3_results}
\input{sections/4_nonAbelian}
\input{sections/5_Conclusions}

%---------------------- Appendices ---------------
\appendix
\clearpage

%Appendix A:
%\input{Supersymmetric_3D_S_duality__COPY_/Supersymmetric_3D_S_duality__COPY_/appendices/N=2components_duality}

\input{appendices/N=2components_duality}

%Appendix B:
%\input{Supersymmetric_3D_S_duality__COPY_/Supersymmetric_3D_S_duality__COPY_/appendices/N=1masterfromN=2master}

\input{appendices/N=1masterfromN=2master}

%Appendix C:
%\input{appendices/nonabelian_interactions}
\input{appendices/nonabelian_interactions}

%\input{appendices/Z: duhamel_formula}
%\input{appendices/Z: total_derivative_cancel}

%------------------ Bibliography --------------
\bibliographystyle{JHEP}
\bibliography{misc/output}

\end{document}

%% file: misc/preamble.tex
\usepackage[utf8]{inputenc}
\usepackage{amsmath,amsfonts,amsthm,amssymb,dsfont}
\usepackage{graphicx,wrapfig,lipsum}
\usepackage[section]{placeins}
\usepackage{cite}
\usepackage{braket}
\usepackage{empheq}
\usepackage{slashed}
\usepackage{tikz}
\usetikzlibrary{shapes.geometric,decorations.markings}
\usepackage{subfigure}
\usepackage[most]{tcolorbox}
\tcbuselibrary{breakable}
\usepackage{bm}
\usepackage[utf8]{inputenc}
\usepackage[T1]{fontenc}

\usepackage[hidelinks]{hyperref}

\hypersetup{
    colorlinks=true,
    linktocpage=true,
    linkcolor=blue,
    urlcolor=blue,
    citecolor=blue,
}

\usepackage{geometry}
\numberwithin{equation}{section}
\usepackage{xcolor}

\newcommand{\chapter}[1]{%
  \clearpage
  \thispagestyle{plain} 
  \section*{#1}         
  \addcontentsline{toc}{section}{#1} 
}

\usepackage{float}

\newcommand{\calN}{\mathcal{N}}

\usepackage{parskip} 

%% file: front/title_page.tex
\begin{titlepage}
\begin{center}

%-------------------- Title ---------------------
$\phantom{.}$\\[2cm]
\noindent{\Large{\textbf{A Master Superspace Action for 3D S-Duality}}}

\vspace{1cm}

%-------------------- Authors ---------------------
Adi Armoni$^{a}$ \footnote{a.armoni@swansea.ac.uk},
Ricardo Stuardo$^{b}$ \footnote{ricardostuardotroncoso@gmail.com}, 
Mark Thomas$^{a}$ \footnote{markdathomas98@gmail.com}

\vspace{0.5cm}

%-------------------- Affiliation ---------------------
\textit{ $^{a}$Department of Physics, Faculty of Science and Engineering\\
        Swansea University, SA2 8PP, UK.  \\ \vspace{0.5cm}
        $^{b}$Departamento de Física, Universidad de Oviedo,
Avda. Federico García Lorca 18, 33007 Oviedo, Spain\\
    and\\
Instituto Universitario de Ciencias y Tecnologías Espaciales de Asturias (ICTEA), Calle de la Independencia 13, 33004 Oviedo, Spain
}

\end{center}

\vspace{0.5cm}
%-------------------- Abstract ---------------------
\centerline{\textbf{Abstract}} 

\vspace{0.5cm}
\input{front/abstract}  

\vspace*{\fill}
\end{titlepage}

%% file: front/abstract.tex
We formulate a `master' partition function in three-dimensional $\mathcal{N}=2$ superspace that for $g^{2}k\neq0$ realises, upon integrating out complementary superfields, both the electric Maxwell--Chern--Simons (MCS) theory and its magnetic $S$-dual: a non-gauge Deser--Jackiw self-dual massive vector times a decoupled level-$k$ Chern--Simons term. The two descriptions share the topological mass $M=\frac{g^{2}k}{2\pi}$ and obey an exact partition-function identity $\mathcal Z_{\rm mag}(g_m^2,k)=\mathcal Z_{\rm ele}(g_e^2,k)$ with $g_e g_m=2\pi$, mapping a weakly coupled MCS theory to a strongly coupled Deser--Jackiw CS theory.
Special cases reproduce the duality between a Chern--Simons theory restricted to flat connections and a Gaiotto--Witten-like theory ($g^{2} = 0$) as well as Maxwell/compact-scalar duality ($k=0$).
We extend the construction to a non-Abelian U$(N)$ gauge group obtaining $\mathcal N=2$ Yang--Mills--Chern--Simons on the electric side and a massive non-gauge vector coupled to level-$k$ Chern--Simons on the magnetic side; the interaction terms between the massive vector and the Chern-Simons term vanish in the Abelian case. Decomposing the $\mathcal N=2$ vector into an $\mathcal N=1$ vector and a real-linear multiplet factorises the master action and yields the $\mathcal N=1$ counterparts. This uplifts the bosonic duality formulated recently to $\mathcal N=2$ and clarifies its non-Abelian and $\mathcal N=1$ reductions.

%% file: sections/1_introduction.tex
\section{Introduction}
Gauge field theories at strong coupling remain a difficult and important problem. In particular, there are few analytic techniques that shed light on the phenomena of confinement and chiral symmetry breaking in QCD-like theories. S-duality, the non-Abelian generalisation of Maxwell's electric-magnetic duality, including its type IIB string theory extension, is a powerful method to handle the strong coupling regime of gauge theories.

Brane configurations, also known as Hanany--Witten configurations \cite{Hanany:1996ie}, lead to a better understanding of both gauge theory and string theory dynamics. In this paper, inspired by type IIB S-duality, we study S-duality in three-dimensional field theories.

In the original paper of Hanany and Witten, the authors considered a brane configuration that consists of a single D3 brane suspended between two parallel NS5 branes and its S-dual: a single D3 brane suspended between two D5 branes. It led to the well-known duality between three-dimensional Maxwell theory and a compact scalar. Our analysis is motivated by this type IIB S-duality action on the underlying fivebrane system. The amount of supersymmetry depends on the relative angle between the fivebranes of the type IIB brane configuration.

Recently, motivated by renewed interest in three-dimensional gauge theories with a Chern--Simons term, \cite{armoni2023s} analysed a brane configuration that includes a D3 brane suspended between an NS5 brane and a tilted $(1,\,k)$ fivebrane. Depending on the relative orientation of the fivebranes, this setup can preserve $\mathcal{N}=2$ supersymmetry, which is the structure we make manifest in our superspace construction below. Such a brane configuration gives rise to a three-dimensional Maxwell--Chern--Simons theory
\begin{equation}
    S_{MCS}\left[A^{(e)}\right] = \int \left( -\frac{1}{2g^{2}}\mathrm{d}A^{(e)}\wedge \star \mathrm{d}A^{(e)} + \frac{k}{4\pi}A^{(e)}\wedge \mathrm{d}A^{(e)} \right) .
\end{equation}
The string theory dual consists of a D3 brane suspended between a D5 brane and a tilted $(-k,1)$ fivebrane. For $g^{2}k\neq 0 $, the resulting field theory on the brane is \cite{armoni2023s}
\begin{equation}
    S_{MCS}^{\text{dual}}\left[A^{(m)},\,b\right] = \int \left(-\frac{g^{2}}{2}A^{(m)}\wedge \star A^{(m)} - \frac{\pi}{k}A^{(m)}\wedge \mathrm{d}A^{(m)} + \frac{k}{4\pi}b\wedge \mathrm{d}b \right) \,.
\end{equation}
This is a Deser--Jackiw massive vector theory \cite{Deser:1984kw} \footnote{The supersymmetric extension of Deser-Jackiw theory was derived in \cite{Karlhede:1986qd}.}, with dynamical field $A^{(m)}$, together with a decoupled topological level-$k$ Chern--Simons sector described by the gauge field $b$ \footnote{The same duality was rederived recently using different arguments in \cite{Chen:2025buv}.}. Both sides exhibit a generated mass
\begin{equation}
    M = \frac{g^{2}k}{2\pi}.
\end{equation}

A natural next step is to investigate how this set of dualities generalises under supersymmetry. In this work, we present a systematic way to embed three-dimensional $\mathcal{N}=2$ supersymmetry into the bosonic MCS master action. Concretely, we generalise the master partition function\footnote{The master field \eqref{Eq: Adi's Master Lagrangian} includes a 1-form $A^{(m)}$ which is {\it not} a gauge field, but a vector. It is possible to replace it by a Stückelberg field $A^{(m)}-\mathrm{d} \phi$, such that now $A^{(m)}$ is a gauge field.}
\begin{equation}
    \label{Eq: Adi's Master Lagrangian}
    \mathcal{Z}_{\text{Master}}^{MCS} = \int \mathcal{D}A^{(m)}\mathcal{D}A^{(e)}\exp i \int \left[-\frac{g^{2}}{2}A^{(m)}\wedge \star A^{(m)} + A^{(m)}\wedge \mathrm{d}A^{(e)} + \frac{k}{4\pi}A^{(e)}\wedge \mathrm{d}A^{(e)}\right]
\end{equation}
so that it accommodates a pair of $\mathcal{N}=2$ supersymmetric multiplets. By enforcing manifest supersymmetry at the level of the master partition function, we ensure that supersymmetry holds on each side of the resulting duality. The construction keeps the topological U$(1)$ sector explicit, and makes all supersymmetry transformations manifest.

For $g^{2}k\neq 0 $, integrating out the auxiliary variables on one side produces an $\mathcal{N}=2$ Maxwell--Chern--Simons theory; integrating the other side yields an $\mathcal{N}=2$ massive vector theory, plus a decoupled level-$k$ Chern--Simons multiplet.

For $g^{2}k\neq 0$, since the Abelian master action is quadratic in both electric and magnetic variables, the path integral is Gaussian. We thus derive the exact, invertible, and gauge-covariant linear map between the electric and magnetic variables. For $g^{2}k=0$, only the gauge-invariant content is fixed. We produce a component-by-component dictionary corresponding to the resulting equations.

After discussing the $\mathcal{N}=2$ master construction, we group components by their $\mathcal{N}=1$ blocks and demonstrate that a smaller $\mathcal{N}=1$ S-duality can be derived from the same master partition function. Following this, we explicitly express the $\mathcal{N}=2$ master field in terms of $\mathcal{N}=1$ superfields in order to derive an $\mathcal{N}=1$ master partition function, from which the corresponding $\mathcal{N}=1$ S-duality can be derived.

We then generalise this analysis to a $U(N)$ gauge group, which corresponds to a string setup with $N$ D3 branes. We promote all Abelian superfields to Lie-algebra valued fields, and introduce a Stückelberg combination that transforms covariantly, becoming gauge-invariant in the Abelian limit. The resulting master partition function is gauge-covariant and collapses correctly in the Abelian limit. As in the Abelian case, for non-zero $g^{2}k$, we integrate out the magnetic variables to obtain the electric theory, and perform a change of variables to obtain the magnetic theory. This results in (non-Abelian) Yang--Mills--Chern--Simons theory on the electric side, extending the result of the Abelian case. On the magnetic side, the change of variables produces a gauge-covariant Deser-Jackiw massive vector with a coupling to a Chern--Simons term. We decompose the result into an Abelian part and a non-Abelian coupling, and show that the non-Abelian coupling vanishes in the Abelian limit. Because the non-Abelian master action is polynomial in gauge-covariant superfields, a covariant superspace variation yields an exact gauge-covariant electric $\longleftrightarrow$ magnetic map that is fully fixed for $g^{2}k\neq 0$.

Finally, we perform the $\mathcal{N}=1$ non-Abelian uplift by promoting the Abelian master partition function to adjoint-valued superfields and carrying out the same analysis as in $\mathcal{N}=2$. This recovers the Yang--Mills-Chern--Simons $\leftrightarrow$ Deser--Jackiw-Chern--Simons duality for non-zero $g^{2}k$. 

In Section \ref{Sec: Conventions} we review our superspace conventions and set up the $\mathcal{N}=2$ multiplets used throughout. Section \ref{Sec: Abelian results} presents the Abelian $\mathcal{N}=2$ master partition function, derives the electric-magnetic map, and analyses the $g^{2}k\neq0$, $g^{2}=0$, and $k=0$ regimes. In Section \ref{Section: Non-Abelian S-duality}, we extend the construction to a non-Abelian $\mathrm{U}(N)$ gauge group, introduce a covariant Stückelberg combination, and obtain the Yang--Mills--Chern--Simons $\leftrightarrow$ Deser--Jackiw--Chern--Simons duality. Finally, Section \ref{Sec: Conclusions} summarises our results and discusses future directions.

%% file: sections/2_conventions.tex
\section{Review of 3D  \texorpdfstring{$\mathcal{N}=2$}{N=2} SUSY}
\label{Sec: Conventions}
Our presentation follows the 3D $\mathcal{N}=2$ formalism obtained by dimensional reduction of 4D $\mathcal{N}=1$ superspace \cite{Wess:1992cp,tong2021supersymmetric}. We work on a boundaryless three-dimensional Minkowski spacetime with mostly-minus metric signature $\eta=\mathrm{diag}\left(+,\,-,\,-\right)$, and take the gamma matrices
\begin{equation}
    \gamma^{0} = \sigma^{2},\qquad \gamma^{1} = i\sigma^{3},\qquad \gamma^{2} = i\sigma^{1},
\end{equation}
where $\sigma^{i}$ are the Pauli matrices. We use the Levi--Civita spacetime normalisation $\varepsilon^{012}=+1$. Both $4D$ $\mathcal{N}=1$ superspace and $3D$ $\mathcal{N}=2$ superspace have four real supercharges. In three dimensions, we organise them as a complex pair, and we accordingly work with complex Grassmann coordinates $\theta_{\alpha}$ and $\bar{\theta}_{\alpha}$ that are normalised such that 
\begin{equation}
    \int \mathrm{d}^{2}\theta\,\theta^{2}=1,\quad \int \mathrm{d}^{2}\bar{\theta}\,\bar{\theta}^{2}=1,
\end{equation}
where $\theta^{2} = \theta^{\alpha}\theta_{\alpha}
$ and $\bar{\theta}^{2} = \bar{\theta}^{\alpha}\bar{\theta}_{\alpha}$. We define the covariant derivatives
\begin{align}
     D_\alpha &= \frac{\partial}{\partial\theta^\alpha}
    + i\,\gamma^m_{\alpha\beta}\,\bar\theta^\beta\,\partial_m,\\
    \bar{D}_\alpha &= -\frac{\partial}{\partial\bar\theta^\alpha}
    - i\,\theta^\beta\,\gamma^m_{\beta\alpha}\,\partial_m,
\end{align}
which satisfy the standard superspace anticommutation relations
\begin{equation}
    \left\{D_{\alpha},\,\bar{D}_{\beta}\right\} = -2i\gamma^{m}_{\alpha\beta}\partial_{m},
\end{equation}
and 
\begin{equation}
    \left\{D_{\alpha},\,D_{\beta}\right\} = \left\{\bar{D}_{\alpha},\,\bar{D}_{\beta}\right\}=0.
\end{equation}
All spinor index contractions are made north--west to south--east, meaning for spinors $\psi$ and $\chi$, 
\begin{equation}
    \psi\chi\equiv \psi^{\alpha}\chi_{\alpha},\quad \bar{\psi}\bar{\chi}\equiv \bar{\psi}^{\alpha}\bar{\chi}_{\alpha},\quad \psi\bar{\chi} = \psi^{\alpha}\bar{\chi}_{\alpha}.
\end{equation}
The underlying construction is based on \cite{aragone1983n}, though we adopt modernised notation closer in spirit to \cite{Intriligator:2013lca}. Chiral multiplets $\Omega$, defined by $\bar{D}_{\alpha}\Omega=0$, contain a complex scalar $\omega$, a complex two-component (Dirac) fermion $\rho$, and a complex auxiliary field $G$. In components this is 
\begin{equation}
    \Omega = \omega + \sqrt{2}\theta\rho + \theta^{2}G+i\theta\gamma^{m}\bar{\theta}\partial_{m}\omega
    + \frac{i}{\sqrt{2}}\theta^{2}\left(\partial_{m}\rho\right)\gamma^{m}\bar{\theta} + \frac{1}{4}\theta^{2}\bar{\theta}^{2}\partial^{2}\omega.
\end{equation}
For the vector multiplet $V=V^{\dagger}$, its component expansion is \cite{aragone1983n, tong2021supersymmetric}
\begin{align}
    \label{Eq: 3D N=2 multiplet}
    V\left(x,\,\theta,\,\bar{\theta}\right) &= C + \theta\chi -\bar{\theta}\bar{\chi}+i\theta^{2}N+i\bar{\theta}^{2}N^{\dagger} + i\theta\gamma^{m}\bar{\theta}A_{m} + \theta\bar{\theta}\phi \nonumber \\&+ \theta^{2}\bar{\theta}\left(\bar{\lambda} + \frac{i}{2}\gamma^{m}\partial_{m}\chi\right) +\bar{\theta}^{2}\theta\left(\lambda - \frac{i}{2}\gamma^{m}\partial_{m}\bar{\chi}\right)\nonumber\\
    &- \frac{1}{2}\theta^{2}\bar{\theta}^{2}\left(D + \frac{1}{2}\partial^{2}C\right).
\end{align}
Here, $C$ and $D$ are real scalars, $N$ is a complex scalar and $\chi_{\alpha}$ and $\lambda_{\alpha}$ are two-component Dirac fermions. The scalar $\phi$ is the real scalar obtained from the dimensional reduction of the $4D$ gauge field. 

We now focus on the Abelian case, leaving the non-Abelian generalisation for Section \ref{Section: Non-Abelian S-duality}. Here, the field strength is in the linear multiplet, defined by
\begin{equation}
    W\left(V\right) = \frac{i}{2}\bar{D}^{\alpha}D_{\alpha}V.
\end{equation}
In the Abelian case, we drop the explicit notation for $V$ dependence, writing $W$ in place of $W(V)$. The field strength satisfies $W=W^{\dagger}$ and $D^{2}W = \bar{D}^{2}W=0$. In terms of the components of the vector multiplet
\begin{align}
    \label{Eq: Abelian W components}
    W &= i\left(\phi - \theta\bar{\lambda} - \bar{\theta}\lambda + \theta\bar{\theta}D +i\,\theta\gamma_{l}\bar{\theta}\varepsilon^{lmn}\partial_{m}A_{n}-\frac{i}{2}\theta^{2}\bar{\theta}\slashed{\partial}\bar{\lambda}+\frac{i}{2}\bar{\theta}^{2}\theta\slashed{\partial}\lambda+\frac{1}{4}\theta^{2}\bar{\theta}^{2}\partial^{2}\phi\right).
\end{align}
Notice that the fields $C$, $\chi$, and $N$ from the vector multiplet $V$ do not appear in $W$. This reflects the gauge freedom of the superfield, whose transformation properties are discussed below. 

Supersymmetric actions are constructed using the superfields above and integrating over the whole of superspace. The Maxwell, Chern-Simons and mass terms are given by
\begin{equation}
    \label{eq:DefinitionLagrangians}
    \mathcal{L}_{\text{Maxwell}} =\frac{1}{g^{2}}\int \mathrm{d}^{4}\theta\, W^{2}, \quad
    \mathcal{L}_{\text{CS}} = \frac{ik}{2\pi}\int \mathrm{d}^{4}\theta\, V\, W, \quad
    \mathcal{L}_{\text{mass}} = m^{2}\int \mathrm{d}^{4}\theta\, V^{2}, 
\end{equation}
where $g$ is the gauge coupling, $k$ the Chern-Simons level and $m$ the mass of the gauge field. In components, each term reads
\begin{align}
        &\int \mathrm{d}^{4}\theta \,W^{2} = \frac{1}{2}\left(\partial_{m}\phi\right)^{2} - i\bar{\lambda}\slashed{\partial}\lambda + \frac{1}{2}D^{2}-\frac{1}{4}F^{mn}F_{mn}, \label{eq:LagrangianMaxwell}\\
        &\int \mathrm{d}^{4}\theta\,VW = -iD\phi + i\bar{\lambda}\lambda + \frac{i}{2}\varepsilon^{lmn}A_{l}\partial_{m}A_{n},\label{eq:LagrangianCS}\\   
    &\begin{aligned}
        \int \mathrm{d}^{4}\theta\,V_{}^{2} &= \frac{1}{2}\left(\partial_{m}C_{}\right)^{2} -C_{}D_{}-2N_{}^{\dagger}N_{} + \frac{1}{2}A_{m}^{}A^{m}_{}-\frac{1}{2}\phi_{}^{2} \\
        &\phantom{=} 
        -\chi_{} \left(\lambda_{}-\frac{i}{2}\slashed{\partial}\bar{\chi}_{}\right)+\bar{\chi}\left(\bar{\lambda}+\frac{i}{2}\slashed{\partial}\chi\right) \label{eq:LagrangianMass}
        \end{aligned}
\end{align}
In addition, we will use two vector superfields $V^{(m)}$ and $V^{(e)}$ of the form \eqref{Eq: 3D N=2 multiplet}. We have
\begin{equation}
    \label{Eq:LagrangianCrossTerm}
    \int \mathrm{d}^{4}\theta\,V^{(m)}W^{(e)} = -\frac{i}{2}\left(\phi^{(m)}D^{(e)}+\phi^{(e)}D^{(m)}\right)+\frac{i}{2}\varepsilon^{lmn}A_{l}^{(m)}\partial_{m}A_{n}^{(e)} + \frac{i}{2}\left(\bar{\lambda}^{(m)}\lambda^{(e)} + \bar{\lambda}^{(e)}\lambda^{(m)}\right),
\end{equation}
with $W^{(e)} = W\left(V^{(e)}\right)$ and, when required, $W^{(m)} = W\left(V^{(m)}\right)$. Actions containing only \eqref{eq:LagrangianMaxwell}, \eqref{eq:LagrangianCS}, and \eqref{Eq:LagrangianCrossTerm} are invariant under the Abelian gauge transformation
    \begin{equation}\label{eq:GaugeTransformation}
        V\rightarrow V + \Lambda^{g}, \quad
        \Lambda^{g} = i(\Omega^{g} -\Omega^{\dagger\,g}),
    \end{equation}
with $\Omega^{g}$ a chiral multiplet. Each vector superfield transforms independently according to \eqref{eq:GaugeTransformation}. Under this transformation $\phi$, $\lambda$ and $D$ are gauge invariant, while 
\begin{align}\label{Eq: Gauge transformation in components}
    C&\rightarrow C -2\, \text{Im}\left(\omega^{g}\right), \quad
    \chi\rightarrow \chi+\sqrt{2}i\rho^{g}, \quad
    N\rightarrow N+G^{g}, \quad
    A_{m}\rightarrow A_{m}-2\,\text{Re}\left(\partial_{m}\omega^{g}\right).
\end{align}
The components of $W$ \eqref{Eq: Abelian W components} assemble into gauge-invariant combinations, so $W$ itself is gauge invariant.  Choosing gauge parameters $\omega^{g},\,\rho^{g},\,G^{g}$ such that $C=\chi=N=0$ produces the Wess-Zumino gauge representative
\begin{equation}
    V_{\text{WZ}} = i\theta\gamma^{m}\bar{\theta}A_{m}+\theta\bar{\theta}\phi + \theta^{2}\bar{\theta}\bar{\lambda}+\bar{\theta}^{2}\theta\lambda - \frac{1}{2}\theta^{2}\bar{\theta}^{2}D.
\end{equation}
On the other hand, an action containing the mass term \eqref{eq:LagrangianMass} is not invariant under the gauge transformation \eqref{eq:GaugeTransformation}. Consequently, the gauge freedom required to impose the Wess-Zumino gauge is lost, and one cannot fix $V$ to $V_{\text{WZ}}$ inside a mass term. 

To restore gauge invariance to this term, we introduce a Stückelberg chiral multiplet $\Omega^{(S)}$, with
\begin{equation}
    \Lambda^{(S)} = i\left(\Omega^{(S)} - \Omega^{\dagger\,(S)}\right),\quad \bar{D}_{\alpha}\Omega^{(S)} = 0,
\end{equation}
and note that the gauge-invariant combination
\begin{equation}
    V' = V-\Lambda^{(S)}
\end{equation}
is invariant under 
\begin{equation}
    V\rightarrow V + \Lambda^{g},\quad \Omega^{(S)}\rightarrow \Omega^{(S)}+\Omega^{g}.
\end{equation}
Thus, to retain a gauge-invariant action, any term that fails to be gauge invariant when written in terms of $V$ must instead be expressed using the gauge-invariant combination $V - \Lambda^{(S)}$.

Since any real superfield may be written as
\begin{equation}
    V = V_{\text{WZ}} + \Lambda,
\end{equation}
and since $V - \Lambda^{(S)}$ is gauge invariant, we are free to use the Wess-Zumino representative of $V$ inside the action and throughout all component calculations. Concretely, we will always evaluate
\begin{equation}
    V - \Lambda^{(S)} = V_{\text{WZ}} - \Lambda^{(S)}.
\end{equation}
This keeps the gauge invariance manifest while ensuring that the physical degrees of freedom remain those of a massive vector multiplet: although the Stückelberg multiplet introduces additional fields, the associated gauge redundancy removes them, so the number of propagating degrees of freedom is unchanged.

With this, we can proceed to the analysis of the master partition function which realises 3D S-duality.

%% file: sections/3_results.tex
\section{A Proposal for 3D \texorpdfstring{$\calN=2$}{N=2} S-Duality}
\label{Sec: Abelian results}
We propose the following  master partition function
\begin{equation}\label{eq:MasterN=2}
    \begin{aligned}
    \mathcal{Z}= \int \mathcal{D}V^{(m)}\mathcal{D}V^{(e)}\mathcal{D}\Lambda^{(S)}\exp \left(i \int \mathrm{d}^{4}\theta\,\int\,\mathrm{d}^{3}x\, \right.&\left[\frac{g^{2}}{\left(2\pi\right)^{2}}\left(V^{(m)}-\Lambda^{(S)}\right)^{2}
    \right.\\&\left.\left.+\frac{{2i}}{2\pi}\,\left(V^{(m)}-\Lambda^{(S)}\right)W^{(e)}
    +\frac{ik}{2\pi}V^{(e)}W^{(e)}\right]\right).
    \end{aligned}
\end{equation}
Here both $V^{(e)}$ and $V^{(m)}$ are dynamical supersymmetric $\mathcal{N}=2$ compact $\mathrm{U}(1)$ vector multiplets in the Wess-Zumino gauge. The field $\Lambda^{(S)}$ is the Stückelberg compensator $\Lambda^{(S)} = i\left(\Omega^{(S)}-\Omega^{\dagger\,(S)}\right)$, with $\Omega^{(S)}$ a chiral $\mathcal{N}=2$ multiplet. The theory depends on two constants, the real $g^{2}\geq 0$ and the integer Chern--Simons level $k$. We exclude the point $\left(g^{2},\,k\right) = \left(0,\,0\right)$, which we do not analyse.

With $\Lambda^{(S)}$ playing the role of a Stückelberg field, this allows us to define the partition function in terms of two gauge vector multiplets $V^{(m)}$ and $V^{(e)}$. Unlike in \eqref{Eq: Adi's Master Lagrangian}, where $A^{(m)}$ is itself gauge invariant, here the action's gauge invariance is restored by the compensator $\Lambda^{(S)}$. In this formulation, the Stückelberg multiplet is interpreted as a magnetic variable.

This partition function is the $\calN=2$ generalization of the one used to obtain S-duality for Maxwell-Chern-Simons theories \cite{armoni2023s}. As in that case,  the ``electric" (``magnetic") theory is obtained from \eqref{eq:MasterN=2} by integrating out the ``magnetic" (``electric") degrees of freedom. For $g^{2}k\neq0$, we find the electric side to be  $\mathcal{N}=2$ Maxwell-Chern-Simons, while the magnetic side is an $\mathcal{N}=2$ version of Deser-Jackiw-Chern-Simons. We explain this below.

\subsection{Electric theory} \label{Subsec: Electric side}
The electric side is obtained by integrating out $V^{(m)}$ and $\Lambda^{(S)}$. For non-zero $g^{2}$, this is a Gaussian integral, yielding the partition function
\begin{equation}
    \label{Eq: Electric side partition function non-zero g}
    \mathcal{Z}^{\text{Electric}}_{g^{2}\neq0}= \int \mathcal{D}V^{(e)}\exp \left(i \,\int\,\mathrm{d}^{3}x\, \int \mathrm{d}^{4}\theta \left[\frac{1}{g^{2}}\left(W^{(e)}\right)^{2} + \frac{ik}{2\pi}V^{(e)} W^{(e)}\right]\right).
\end{equation}
Using \eqref{eq:DefinitionLagrangians}, \eqref{Eq: Electric side partition function non-zero g} is the Maxwell--Chern--Simons action with  coupling $g$ and level $k$. Integrating out the auxiliary field $D^{(e)}$ gives
\begin{equation}
    D^{(e)}=-M\phi^{(e)},
\end{equation}
where
\begin{equation}
     M= \frac{g^{2}k}{2\pi}.
\end{equation}
The action in components is then
\begin{equation}
\begin{aligned}
    S^{(e)}_{g^{2}\neq0} = \frac{1}{g^{2}} \int \mathrm{d}^{3}x\;&\left[\frac{1}{2}\left(\partial_{m}\phi^{(e)}\right)^{2} - \frac{1}{2}M^{2}\phi_{(e)}^{2}-\bar{\lambda}^{(e)}\left[i\slashed{\partial} +M\right]\lambda^{(e)}\right.\nonumber\\&\left.-\frac{1}{4}F_{(e)}^{mn}F^{(e)}_{mn} - \frac{1}{2}M\varepsilon^{mnp}A^{(e)}_{p}\partial_{m}A^{(e)}_{n}\right]. 
\end{aligned}
\end{equation}
All propagating components sit in the single topologically massive $\mathcal{N}=2$ vector multiplet $V^{(e)}$ of topological mass $M$; the mass originates from the Chern--Simons term and it preserves gauge invariance.

In the case $g^{2}=0$, the integral over $V^{(m)}$ and $\Lambda^{(S)}$ is no longer Gaussian. Instead, the path integral over $V^{(m)}$ imposes a flatness constraint on $W^{(e)}$, leaving:
\begin{equation}
    \mathcal{Z}^{\text{Electric}}_{g^{2}=0}= \int \mathcal{D}V^{(e)}\,\delta\left[W^{(e)}\right]\exp \left(i \int\,\mathrm{d}^{3}x\,\int \mathrm{d}^{4}\theta\,\left[ \,\frac{ik}{2\pi}V^{(e)} W^{(e)}\right]\right).
\end{equation}
The delta functional restricts the path integral to configurations satisfying $W^{(e)}=0$, yielding a constrained level-$k$ $\mathcal{N}=2$ Chern--Simons theory rather than ordinary level-$k$ Chern--Simons theory.

\subsection{Magnetic theory}
\label{Subsec: Magnetic side}
We now proceed to the study of the magnetic side of the duality, for which we need to integrate out $V^{(e)}$ in \eqref{eq:MasterN=2}. In contrast to the electric side, the case $k\neq0$ must be treated separately from the $k=0$ point; we explain this below.

In the case of non-zero $k$, we may make the change of variables
\begin{equation}
    \label{Eq: Supersymmetric change of variables}
    V^{(e)} = B - \frac{1}{k}\left(V^{(m)}-\Lambda^{(S)}\right).
\end{equation}
Here $B$ is a dynamical $\mathrm{U}(1)$ gauge superfield in the Wess-Zumino gauge. Upon this change, the master partition function becomes 
\begin{align}
\label{Eq: Magnetic side partition function}
    \mathcal{Z}= \int \mathcal{D}V^{(m)}\mathcal{D}\Lambda^{(S)}\,\mathcal{D}B\,&\exp \left(i \int \mathrm{d}^{4}\theta\,\,\int\,\mathrm{d}^{3}x\,  \left[\frac{g^{2}}{\left(2\pi\right)^{2}}\left(V^{(m)}-\Lambda^{(S)}\right)^{2} \right.\right.\nonumber\\&\left.\left.\qquad\qquad-\frac{i}{2\pi k}\left(V^{(m)}-\Lambda^{(S)}\right)W^{(m)} +\frac{ik}{2\pi} B  W^{(B)}\right]\right),
\end{align}
where $W^{(B)}=W(B)$ is the field strength of $B$. For $g^{2}k\neq 0$, this is the theory of a linearly supersymmetric non-gauge $\mathcal{N}=2$ massive Deser--Jackiw vector alongside a decoupled $\mathcal{N}=2$ level-$k$ Chern--Simons theory. The decoupling means we may write the factorisation
\begin{equation}
    \mathcal{Z}=\mathcal{Z}_{\text{DJ}}^{\mathcal{N}=2}\times\mathcal{Z}_{\text{CS}(k)}^{\mathcal{N}=2}.
\end{equation}
Expanding \eqref{Eq: Magnetic side partition function} in terms of components, we have the action
\begin{align}
    \label{Eq: Magnetic action linear supersymmetry}
    S_{\text{Magnetic}}^{k\neq 0}&= \int\mathrm{d}^{3}x\,\Bigg[
    \frac{g^{2}}{\left(2\pi\right)^{2}}\left\{\frac{1}{2}\left(\partial_{m}C^{(S)}\right)^{2} +C^{(S)}D^{(m)}-N^{\dagger\,(S)}N^{(S)}\right.\nonumber\\ &\left. + \frac{1}{2}\left(A_{m}^{(m)}-A_{m}^{(S)}\right)\left(A^{m}_{(m)}-A^{m}_{(S)}\right)-\frac{1}{2}\phi_{(m)}^{2}\right.\nonumber\\
    &\left.+\chi^{(S)}
    \left(\lambda^{(m)}+\frac{i}{2}\slashed{\partial}\bar{\chi}^{(S)}\right)-\bar{\chi}^{(S)}\left(\bar{\lambda}^{(m)}-\frac{i}{2}\slashed{\partial}\chi^{(S)}\right)\right\}\nonumber\\
    &-\frac{i}{2\pi k }\left\{-iD^{(m)}\phi^{(m)} + i\bar{\lambda}^{(m)}\lambda^{(m)} + \frac{i}{2}\varepsilon^{lmn}\left(A^{(m)}_{l}-A_{l}^{(S)}\right)\partial_{m}\left(A^{(m)}_{n}-A_{n}^{(S)}\right)\right\}\nonumber\\
    &+\frac{ik}{2\pi}\left\{-iD^{(B)}\phi^{(B)} + i\bar{\lambda}^{(B)}\lambda^{(B)} + \frac{i}{2}\varepsilon^{lmn}A^{(B)}_{l}\partial_{m}A^{(B)}_{n}\right\}\Bigg].
\end{align}
For $g^{2}k\neq 0$, the scalar $N^{(S)}$ is auxiliary and integration removes it trivially. Integration over $D^{(m)}$ requires
\begin{equation}
    C^{(S)} = \frac{1}{M}\phi^{(m)},
\end{equation}
and similarly integration over $\lambda^{(m)}$ and $\bar{\lambda}^{(m)}$ imposes
\begin{equation}
    \bar{\lambda}^{(m)} = -M\chi^{(S)}
\end{equation}
and
\begin{equation}
    \lambda^{(m)} = M\bar{\chi}^{(S)}. 
\end{equation}
These integrals leave the magnetic side action as
\begin{align}
    \label{Eq: Magnetic action finite mass}
    S_{\text{Magnetic}}^{g^{2}k\neq0} = &\int\mathrm{d}^{3}x\,\Bigg[ \frac{1}{g^{2}k^{2}}\Bigg\{\frac{1}{2}\left(\partial_{m}\phi^{(m)}\right)^{2} - \frac{1}{2}M^{2}\phi_{(m)}^{2} \Bigg\}\nonumber\\&+ \frac{1}{2\pi k }\Bigg\{\frac{M}{2}\left(A^{(m)}_{m}-A^{(S)}_{m}\right)\left(A_{(m)}^{m}-A_{(S)}^{m}\right) \nonumber \\ &+ \frac{1}{2}\varepsilon^{lmn}\left(A_{l}^{(m)} - A_{l}^{(S)}\right)\partial_{m}\left(A_{n}^{(m)}-A_{n}^{(S)}\right) - M\bar{\chi}^{(S)}\left(i\slashed{\partial}-M\right)\chi^{(S)}\Bigg\}\nonumber\\
    &+\frac{ik}{2\pi}\left\{-iD^{(B)}\phi^{(B)} + i\bar{\lambda}^{(B)}\lambda^{(B)} + \frac{i}{2}\varepsilon^{lmn}A^{(B)}_{l}\partial_{m}A^{(B)}_{n}\right\}\Bigg].
\end{align}
This is the $\mathcal{N}=2$ supersymmetric Deser--Jackiw vector of mass $M$ alongside a decoupled $\mathcal{N}=2$ level-$k$ Chern--Simons theory.

At the strict point $g^2=0$, with $k\neq0$, we remove the first term in \eqref{eq:MasterN=2} without making a change of variables. Using that $W\left(\Lambda^{(S)}\right)=0$, the Stückelberg multiplet then decouples from the remaining superspace action, so its path integral contributes only an overall normalisation. We are left with
\begin{align}
    \label{Eq: Supersymmetric GW}
    \mathcal{Z}_{g^{2}=0}= \int \mathcal{D}V^{(m)}\mathcal{D}V^{(e)}\,\exp \left(i \int \mathrm{d}^{4}\theta\,\int\,\mathrm{d}^{3}x\, \left[\frac{2
    i}{2\pi}\,V^{(m)}W^{(e)}+\frac{ik}{2\pi}V^{(e)}W^{(e)}\right]\right).
\end{align}
This is a standalone $\mathrm{U}(1)^{2}$ BF--Chern--Simons theory, which is what we call an $\mathcal{N}=2$ supersymmetric Gaiotto--Witten-like theory\footnote{By Gaiotto--Witten-like we mean that the BF--Chern--Simons interaction structure is analogous to that appearing in section 8.3 of \cite{Gaiotto:2008ak}; in our construction both $\mathrm{U}(1)$ vector multiplets are dynamical three-dimensional fields.}. This is dual to the constrained $\mathcal{N}=2$ level-$k$ Chern--Simons theory obtained on the electric side; integration over $V^{(m)}$ imposes $W^{(e)}=0$. 

Approaching $g^{2}=0$ from $g^{2}>0$, the limit $g^{2}\to 0$ sends the common topological mass $M$ to zero on both sides of the duality. The mass gap closes, and so this limit need not commute with the infrared limit. We therefore distinguish the strict $g^{2}=0$ theory from the $g^{2}\to 0$ limit. 

In the case of $k=0$, we may directly integrate over $V^{(e)}$. Because $W\left(\Lambda^{(S)}\right)=0$, this results in a magnetic side action
\begin{equation}
    \mathcal{Z}^{\text{Magnetic}}_{k=0}= \int \mathcal{D}V^{(m)}\mathcal{D}\Lambda^{(S)}\,\delta\left[W\left(V^{(m)}\right)\right]\exp\left( i \int\,\mathrm{d}^{3}x\,\int \mathrm{d}^{4}\theta\,\left[\frac{g^{2}}{\left(2\pi\right)^{2}}\left(V^{(m)}-\Lambda^{(S)}\right)^{2}\right]\right).
\end{equation}
On each topological sector $\mathfrak{h}\in H^{1}\left(\mathcal{M},\,\mathrm{U(1)}\right)$, a flat representative $V_{\text{harm}}^{\mathfrak{h}}$ may be chosen; any flat $V^{(m)}$ may be written
\begin{equation}
    V_{(m)}^{\mathfrak{h}} = i\left(\Sigma-\Sigma^{\dagger}\right)+V_{\text{harm}}^{\mathfrak{h}} ;\quad W\left(V_{\text{harm}}^{\mathfrak{h}}\right)=0;\quad \bar{D}_{\alpha}\Sigma=0.
\end{equation}
Using the Stückelberg redundancy in $V'^{(m)} = V^{(m)}-\Lambda^{(S)}$, the exact part is absorbed into $\Lambda^{(S)}$, leaving
\begin{equation}
    V^{(m)}-\Lambda^{(S)} = V_{\text{harm}}-i\left(\Omega^{(S)}- \Omega^{\dagger\,(S)}\right),
\end{equation}
with
\begin{equation}
    \int\mathcal{D}V_{\text{harm}} = \sum_{\mathfrak{h}\in H^{1}\left(\mathcal{M},\,\mathrm{U(1)}\right)}\int \mathcal{D}V_{\text{harm}}^{\mathfrak{h}}.
\end{equation}
Summing over harmonic representatives then gives
\begin{equation}
    \mathcal{Z}_{\text{Magnetic}}^{k=0} = \int \mathcal{D}V_{\text{harm}}\,\mathcal{D}\Omega^{(S)}\,\mathcal{D}\Omega^{\dagger\,(S)} \exp \left(i \int \mathrm{d}^{3}x\,\int \mathrm{d}^{4}\theta\,\frac{g^{2}}{\left(2\pi\right)^{2}}\left(V_{\text{harm}}-i\left(\Omega^{(S)}- \Omega^{\dagger\,(S)}\right)\right)^{2}\right).
\end{equation}
For each $\mathfrak{h}$, the mixed term between $V_{\text{harm}}^{\mathfrak{h}}$ and $\Omega^{(S)}$ is a total superspace derivative, so
\begin{equation}
    \int \mathrm{d}^{4}\theta\,\left(V^{(m)}-\Lambda^{(S)}\right)^{2} = \int \mathrm{d}^{4}\theta\,\left[\left(V_{
    \text{harm}}^{\mathfrak{h}}\right)^{2}+2\Omega^{\dagger\,(S)}\Omega^{(S)}\right].
\end{equation}
Therefore the path integral factorises into a purely topological factor
\begin{equation}
    \label{Eq: Topological normalisation}
    \mathcal{Z}_{\text{M}}^{\text{T}} = \sum_{\mathfrak{h}\in H^{1}\left(\mathcal{M},\,\mathrm{U(1)}\right)}\int \mathcal{D}V_{\text{harm}}^{\mathfrak{h}}\exp\left(i\frac{g^{2}}{\left(2\pi\right)^{2}}\int \mathrm{d}^{3}x\,\int\mathrm{d}^{4}\theta\,\left(V_{\text{harm}}^{\mathfrak{h}}\right)^{2}\right),
\end{equation}
capturing the full $\mathrm{U(1)}$ holonomy data, times a free chiral multiplet
\begin{equation}
    \mathcal Z_{\text{Magnetic}}^{k=0}
    =\mathcal{Z}_{\text{M}}^{\text{T}}\,\int \mathcal D\Omega^{(S)}\,\mathcal D\Omega^{\dagger\,(S)}\exp \,\left(i\frac{2g^2}{(2\pi)^2}\int d^3x\,d^4\theta\;\Omega^{\dagger\,(S)}\Omega^{(S)}\right).
\;
\end{equation}
Locally, this is a free $\mathcal{N}=2$ chiral multiplet $\Omega^{(S)}$ (the $\mathcal{N}=2$ supersymmetric dual photon); globally, the theory sums over all flat $\mathrm{U(1)}$ connections. In components, we obtain
\begin{equation}
    S_{\text{Magnetic}}^{k=0} = \int \mathrm{d}^{3}x\ \frac{g^{2}}{\left(2\pi\right)^{2}}\left[\frac{1}{2}\left(\partial_{m}C^{(S)}\right)^{2} + \frac{1}{2}\left(\partial_{m}\alpha^{(S)}\right)^{2} + i\bar{\chi}^{(S)}\slashed{\partial}{\chi}^{(S)}\right],
\end{equation}
where $\alpha^{(S)}$ is a real scalar; it is the dual photon coming from $\Omega^{(S)}$. This is the theory of a free $\mathcal{N}=2$ massless complex scalar. Since the Stückelberg multiplet is $\mathrm{U}(1)$-valued, its group-valued nature forces the dual scalar $\alpha^{(S)}$ to be compact; this will become essential in the discussion of global structure. The duality between the $\mathcal{N}=2$ vector (electric theory) and a dual $\mathcal{N}=2$ scalar is well known in field theory and string theory, and recovering it from the master action \eqref{eq:MasterN=2} provides favorable evidence for our proposal.

\subsection{Statement of the duality}
\setlength{\arrayrulewidth}{2pt}
\begin{table}
\centering
\resizebox{\textwidth}{!}{%
  \renewcommand{\arraystretch}{2.0}%
  \begin{tabular}{|c|c|c|}
    \hline
    \multicolumn{3}{|c|}{%
      \raisebox{1mm}{\(\displaystyle 
        \mathcal{Z} 
        = \int \mathcal{D}V^{(m)} \,\mathcal{D}V^{(e)} \,\mathcal{D}\Lambda^{(S)} \,\exp\,
        i\int \mathrm{d}^{4}\theta\int \mathrm{d}^{3}x \,\left[
          \frac{g^2}{\left(2\pi\right)^{2}}V_{(m)}'^2 
          \;+\;\frac{2i}{2\pi}\,V'^{(m)}W^{(e)} 
          \;+\;\frac{ik}{2\pi}\,V^{(e)}\,W^{(e)}
        \right]
      \)}
    } \\
    \hline
    \textbf{Case} & \textbf{Electric Side} & \textbf{Magnetic Side} \\
    \hline
    \(\displaystyle g^2 \neq 0,\; k \neq 0\)
    & \(\displaystyle
      \begin{array}{@{}c@{}}
        S=\int \mathrm{d}^{4}\theta\,\int\,\mathrm{d}^{3}x\,\left[\frac{1}{g^2}\,W_{(e)}^2 \;+\; \,\frac{ik}{2\pi}\,V^{(e)}\,W^{(e)}\right] 
        \\[-10pt]
        \rule{6cm}{0.5pt}
        \\[-10pt]
        \mathcal{Z}=\mathcal{Z}_{\text{MCS}(k)}^{\mathcal{N}=2}
      \end{array}
    \)
    & \(\displaystyle
      \begin{array}{@{}c@{}}
        S=\int\mathrm{d}^{4}\theta\,\int\,\mathrm{d}^{3}x\,\left[\,\frac{g^2}{\left(2\pi\right)^{2}}\,V_{(m)}'^2
        \;-\; \frac{i}{2\pi k}\,V'^{(m)}\,W'^{(m)}
        \;+\; \,\frac{ik}{2\pi}\,B\,W^{(B)}\right]
        \\[-10pt]
        \rule{12cm}{0.5pt}
        \\[-10pt]
        \mathcal{Z}=\mathcal{Z}_{\text{DJ}}^{\mathcal{N}=2}\times\mathcal{Z}_{\text{CS}(k)}^{\mathcal{N}=2}
      \end{array}
    \) \\
    \hline
    \(\displaystyle g^2 =0,\; k \neq 0\)
    & \(\displaystyle
      \begin{array}{@{}c@{}}
        S=\int \mathrm{d}^{4}\theta\,\int\,\mathrm{d}^{3}x\left[\,\frac{ik}{2\pi}\,V^{(e)}\,W^{(e)}\right]
        \\[-10pt]
        \rule{6cm}{0.5pt}
        \\[-10pt]
        \mathcal{Z} = \mathcal{Z}_{\text{CS}(k)}^{\mathcal{N}=2}\big|_{W^{(e)}=0}
      \end{array}
    \)
    & \(\displaystyle
      \begin{array}{@{}c@{}}
        S= 
        \int \mathrm{d}^{4}\theta\,\int\,\mathrm{d}^{3}x\, \left[\frac{2
    i}{2\pi}\,V^{(m)}W^{(e)}+\,\frac{ik}{2\pi}V^{(e)}W^{(e)}\right]
        \\[-10pt]
        \rule{12cm}{0.5pt}
        \\[-10pt]
        \mathcal{Z}=\mathcal{Z}_{\text{GW}(k)}^{\mathcal{N}=2}
      \end{array}
    \)\\
    \hline
    \(\displaystyle k = 0,\; g^2 \neq 0\)
    & \(\displaystyle
      \begin{array}{@{}c@{}}
        S=\int\mathrm{d}^{4}\theta\,\int\,\mathrm{d}^{3}x\,\left[\frac{1}{g^2}\,W_{(e)}^2\right]
        \\[-10pt]
        \rule{6cm}{0.5pt}
        \\[-10pt]
        \mathcal{Z}=\mathcal{Z}_{\text{M}}^{\mathcal{N}=2}
      \end{array}
    \)
    & \(\displaystyle
      \begin{array}{@{}c@{}}
        S=\int\mathrm{d}^{4}\theta\,\int\,\mathrm{d}^{3}x\,\left[\,\frac{2g^2}{\left(2\pi\right)^{2}}\,\Omega^{\dagger\,(S)}\Omega^{(S)}\right]
        \\[-10pt]
        \rule{12cm}{0.5pt}
        \\[-10pt]
        \mathcal{Z} = \mathcal{Z}_{M}^{\mathrm{T}}\mathcal{Z}_{\Omega^{(S)}}^{\mathcal{N}=2}
      \end{array}
    \) \\
    \hline
  \end{tabular} 
} 
\caption{\textit{Summary of the master superspace partition function (top row) and its reduction 
into electric and magnetic actions in the regimes considered. Here $V'^{(m)}= V^{(m)} -\Lambda^{(S)}$, and $W{(\cdot)}=\frac{i}{2}\bar{D}D\left(\cdot\right)$. $S$ denotes the non-trivial part of the bulk action in each cell. $V^{(m)}$ may be evaluated in terms of components as $V^{(m)}_{WZ}$ for all calculations. $B$ is a genuine gauge vector superfield. $\mathcal{Z}_{M}^{\mathrm{T}}$ denotes the normalisation \eqref{Eq: Topological normalisation}.}}
\label{tab:ElectricMagneticSummaryWithMaster}
\end{table}
Table \ref{tab:ElectricMagneticSummaryWithMaster} compiles the outcomes. The electric theory follows by integrating out $V^{(m)}$ and $\Lambda^{(S)}$, giving $\mathcal{N}=2$ Maxwell--Chern--Simons at level $k$ with topological mass $M$, with $M = \frac{g^{2}k}{2\pi}$ for $g^{2}> 0$ and $k\in \mathbb{Z}$, excluding $\left(g^{2},\,k\right)=\left(0,\,0\right)$. On the magnetic side, for $g^{2}k\neq 0$, setting $V^{(e)} = B - \frac{1}{k}\left(V^{(m)}-\Lambda^{(S)}\right)$ with $B$ a genuine $\mathrm{U(1)}$ vector superfield factorises the master partition function into a linearly supersymmetric massive Deser--Jackiw sector of topological mass $M$ times a decoupled $\mathcal{N}=2$ level $k$ Chern--Simons term. For $k=0$, integrating out $V^{(e)}$ imposes $\delta\left[W(V^{(m)} - \Lambda^{(S)})\right]$, and writing $V^{(m)}-\Lambda^{(S)}=V_{\text{harm}}^{\mathfrak{h}}-i\left(\Omega^{(S)}-\Omega^{\dagger\,(S)}\right)$ gives a factorisation into a topological normalisation $\mathcal{Z}_{M}^{\mathrm{T}}$ times a single free chiral Gaussian in $\Omega^{(S)}$. In all cases, the partition functions coincide with common topological mass $M$ (vanishing at $g^{2}k=0$), and the gauge structure matches. For $k\neq0$, large gauge transformations act non-trivially and identically on both sides, while for $k=0$, the action is invariant under large gauge transformations; nevertheless the same global flat-connection sectors exist on both sides. This is explicit in the factorisation of $\mathcal{Z}_{M}^{\text{T}}$ on the magnetic side, and as Maxwell zero modes on the electric, and so any overall topological factor is common. 

For finite $g^{2}>0$, we identify the electric and magnetic couplings from the symmetric quadratic terms in $W^{(e)}$ and $V^{(m)}-\Lambda^{(S)}$: the coefficients are $1/g_{e}^{2}$ and $1/g_{m}^{2}$, which fix the residues of the two-point functions. Accordingly, 
\begin{equation}
    g_{e} = g
\end{equation}
and
\begin{equation}
    g_{m} = \frac{2\pi}{g}.
\end{equation}
The partition functions of the electric and magnetic sides are therefore related via
\begin{equation}
\mathcal Z_{\text{magnetic}}\bigl(g_{m}^{2},k;\,\mathcal{M}\bigr)
   =\mathcal Z_{\text{electric}}\bigl(g_{e}^{2},k\,;\mathcal{M}\bigr).
\end{equation}
Thus the duality is S-duality, with inverted couplings, 
\begin{equation}
    g_{e}g_{m}=2\pi,
\end{equation}
with small $g_{e}$ corresponding to large $g_{m}$, and vice versa.

Varying the master partition function, we have
\begin{equation}
    \label{Eq: Electric to non-gauge magnetic map}
   g^{2}\left(V^{(m)}-\Lambda^{(S)}\right)+2\pi iW^{(e)} = 0
\end{equation}
due to the variation of $V^{(m)}-\Lambda^{(S)}$, and
\begin{equation}
    \label{Eq: Electric to magnetic field strength map}
    W^{(m)}+kW^{(e)} = 0
\end{equation}
due to the variation of $V^{(e)}$. Substituting these relations into each other, we obtain the quantum equations of motion 
\begin{equation}
    \left(\frac{1}{2}\bar{D}D + M\right)\left(V^{(m)}-\Lambda^{(S)}\right)=0
\end{equation}
for the magnetic side, and 
\begin{equation}
    \left(\frac{1}{2}\bar{D}D + M\right)W^{(e)}=0
\end{equation}
on the electric side. In both equations, any operators for which expectation values are calculated must be independent of both $V^{(m)}-\Lambda^{(S)}$ and $V^{(e)}$.

In the case of $M\neq 0$, the quadratic form of the master action is non-degenerate, so integrating out either side produces an invertible map between the correlators of the two descriptions. The change of variables \eqref{Eq: Supersymmetric change of variables} acts as a map from the magnetic side to the electric side. These identities \eqref{Eq: Electric to non-gauge magnetic map} and \eqref{Eq: Electric to magnetic field strength map} can be combined to provide the mapping from the electric to the magnetic side; 
\begin{equation}
     V^{(m)}-\Lambda^{(S)}  = - \frac{2\pi i }{g^{2}} W^{(e)},
\end{equation}
for insertions independent of $V^{(m)}-\Lambda^{(S)}$. Likewise, 
\begin{equation}
    B=  V^{(e)} - \frac{i}{M}W^{(e)}
\end{equation}
for insertions independent of both $V^{(m)}-\Lambda^{(S)}$ and $V^{(e)}$ completes the map from the electric to magnetic side. The component level dictionary of this is provided in appendix \ref{Appendix: N=2 Schwinger Dyson}.

In the massless case, the quadratic form in the master action degenerates and the derived relations impose flatness rather than providing an invertible map. Specifically, for $k=0$, \eqref{Eq: Electric to magnetic field strength map} enforces $ W^{(m)}=0$, imposing flatness on $V^{(m)}$. Following this, \eqref{Eq: Electric to non-gauge magnetic map} then relates the gauge-invariant correlators on the electric side built from the flat magnetic configurations, and vice versa. Similarly, for $g^{2}=0$, \eqref{Eq: Electric to non-gauge magnetic map} imposes flatness on the electric side;  $ W^{(e)}=0$, and then \eqref{Eq: Electric to magnetic field strength map} imposes magnetic side flatness, $ W^{(m)}=0$, with no mapping between the two sides provided. 

\subsection{Reduction to \texorpdfstring{$\mathcal{N}=1$}{N=1}}
We decompose the $\mathcal{N}=2$ vectors as
\begin{equation}
    V\left(x,\,\theta,\,\bar{\theta}\right) = V_{0}+\bar{\theta}^{\alpha}\Gamma_{\alpha} + \bar{\theta}^{2}S
\end{equation}
These components are obtained from $V$ via
\begin{align}
\label{Eq:V0-define}
V_{0}(x,\theta)
&= V(x,\theta,\bar\theta)\Big|_{\bar\theta=0},
\\
\Gamma_\alpha(x,\theta)&=\frac{\partial V}{\partial \bar\theta^\alpha}\Big|_{\bar\theta=0},\\
\label{Eq:S-define}
S(x,\theta)
&= \frac{1}{4}\frac{\partial^{2}V}{\partial\bar{\theta}^{\alpha}\partial \bar\theta_\alpha}
   \Big|_{\bar\theta=0},
\end{align}
which evaluate to
\begin{align}
    V_{0}&=C + \theta\chi+ i\theta^{2}N,\\
    \Gamma_{\alpha}&=-\bar{\chi}_{\alpha} - i\theta^{\beta}\gamma_{\beta\alpha}^{m}A_{m} + 
    \theta_{\alpha}\phi + \theta^{2}\left(\bar{\lambda} + \frac{i}{2}\slashed{\partial}\chi\right)_{\alpha}\\
    S&=iN^{\dagger} + \theta\left(\lambda - \frac{i}{2}\slashed{\partial}\bar{\chi}\right) - \frac{1}{2}\theta^{2}\left(D + \frac{1}{2}\partial^{2}C\right).
\end{align}
We isolate the two $\mathcal{N}=1$ field-strength blocks:
\begin{equation}
    \Sigma = \frac{1}{2} D^\alpha \Gamma_\alpha = iW\big|_{\bar\theta=0},
    \qquad
    W_\alpha = \frac{1}{2} D^\beta D_\alpha \Gamma_\beta,
\end{equation}
where $W$ denotes the $\mathcal{N}=2$ field strength, and $W^{\alpha}$ the $\mathcal{N}=1$ one. The prepotential redundancy $\Gamma_\alpha\rightarrow \Gamma_\alpha+D_\alpha K$, with $K$ an arbitrary real $\mathcal{N}=1$ superfield, implies
\begin{equation}
    W_{\alpha} \rightarrow W_{\alpha},\quad \Sigma \rightarrow \Sigma+\frac{1}{2}D^{2}K.
\end{equation}
We work in an $\mathcal{N}=1$ Wess--Zumino-like gauge that places $\bar{\chi}$ and $\phi$ inside $\Sigma$ so that the $\mathcal{N}=1$ vector block depends only on $W_{\alpha}$, and the scalar block depends only on $\Sigma$. This gauge choice uses only the $\mathcal{N}=1$ prepotential redundancy $\Gamma_{\alpha}\rightarrow \Gamma_{\alpha}+D_{\alpha}K$ and does not invoke the full $\mathcal{N}=2$ gauge symmetry. In particular, writing
\begin{equation}
    \Lambda = \Lambda_{(0)}\left(x,\,\theta\right)+\bar{\theta}^{\alpha}\Lambda_{(1)\alpha}\left(x,\,\theta\right)+\bar{\theta}^{2}\Lambda_{(2)},
\end{equation}
we have
\begin{equation}
    \delta V_{0} = \Lambda_{(0)},\quad \delta \Gamma_{\alpha} = \Lambda_{(1)\alpha},\quad \delta S = \Lambda_{(2)},\quad \delta \Sigma = \frac{1}{2}D^{\alpha}\Lambda_{(1)\alpha},
\end{equation}
so $\Sigma$ is not gauge invariant by itself. However, in the master action, the scalar block, built from $\left(\Sigma;\;V_{0},\,S\right)$, is closed under this symmetry and its variation reduces to a total superspace derivative; equivalently, the scalar contribution to each term in the master action is gauge-invariant up to boundary terms, with the boundary pieces cancelling in the Chern-Simons sector. We will therefore drop such boundary terms. Thus, we may treat the vector block as a functional of $W_{\alpha}$ and the real scalar block as a functional of $\Sigma$, and then integrate out either of the $\mathcal{N}=1$ vector side or the scalar side in order to obtain the complementary theory.

Explicitly, to reduce to $\mathcal{N}=1$ superspace in this notation, we move to chiral coordinates $y^{m} = x^{m} + i\theta\gamma^{m}\bar{\theta}$. With this shift, we find
\begin{align}
    \label{Eq: Chiral coordinates integral trick}
    \int \mathrm{d}^{4}\theta\,F\left(y,\,\theta,\,\bar{\theta}\right) &=-\frac{1}{4}\int \mathrm{d}^{2}\theta\,\bar{D}^{2}F|_{\bar{\theta}=0}.
\end{align}
Also,
\begin{equation}
    \Sigma\left(y,\,\theta\right) = -\phi + \theta\bar{\lambda}
\end{equation}
and 
\begin{equation}
    W_{\alpha}(y,\,\theta)=-\bar{\lambda}_{\alpha}+\theta_{\alpha}D -\frac{1}{2}\left(\gamma^{m}\theta\right)_{\alpha}\varepsilon^{mnp}F_{np}+\frac{i}{2}{\theta}^{2}\left(\slashed{\partial}\bar{\lambda}\right)_{\alpha}.
\end{equation}
Following this (see Appendix \ref{Appendix: N=1 from N=2}), the mass term becomes
\begin{equation}
    \int \mathrm{d}^{4}\theta \left(V^{(m)}-\Lambda^{(S)}\right)^{2} = \int \mathrm{d}^{2}\theta\,\left[2\left(V_{0}^{(m)}-\Lambda_{(0)}^{(S)}\right)\left(S^{(m)}-\Lambda_{(2)}^{(S)}\right) - \frac{1}{2}\left(\Gamma^{(m)}-\Lambda_{(1)}^{(S)}\right)^{2}\right].
\end{equation}
The mixed term becomes
\begin{align}
    \int \mathrm{d}^{4}\theta\,\left(V^{(m)}-\Lambda^{(S)}\right)W^{(e)}&=\int \mathrm{d}^{2}\theta\,\left[-i\left(S^{(m)}-\Lambda_{(2)}^{(S)}\right)\Sigma^{(e)} + \frac{i}{4}\left(\Gamma^{(m)\,\alpha}-\Lambda_{(1)}^{(S)\,\alpha}\right)W_{\alpha}^{(e)}\right]
\end{align}
up to total superspace derivatives, and the Chern-Simons term becomes
\begin{align}
    \int \mathrm{d}^{4}\theta\,V^{(e)}W^{(e)}=&\int \mathrm{d}^{2}\theta\,\left[-iS^{(e)}\Sigma^{(e)} + \frac{i}{4}\Gamma^{(e)\,\alpha}W_{\alpha}^{(e)}\right].
\end{align}
These expressions lead to a decomposition into $\mathcal{N}=1$ vector and $\mathcal{N}=1$ scalar contributions. The vector master partition function is
\begin{align}
    \mathcal{Z}_{\mathcal{N}=1}^{\text{Vector}} &=\int \mathcal{D}\Gamma^{(m)}\mathcal{D}\Gamma^{(e)}\mathcal{D}\Lambda_{(1)}^{(S)}\exp\left( i \,S_{\mathcal{N}=1}^{\text{Vector}}\right),
\end{align}
where
\begin{align}
    S_{\mathcal{N}=1}^{\text{Vector}}= \int \mathrm{d}^{2}\theta\int \mathrm{d}^{3}x\,&\left[-\frac{g^{2}}{2\left(2\pi\right)^{2}}\left(\Gamma^{(m)\,\alpha}-\Lambda_{(1)}^{(S)\,\alpha}\right)\left(\Gamma^{(m)}_{\alpha}-\Lambda_{(1)\,\alpha}^{(S)}\right)\right.\nonumber\\&\left. - \frac{1}{4\pi}\left(\Gamma^{(m)\,\alpha}-\Lambda_{(1)}^{(S)\,\alpha}\right)W_{\alpha}^{(e)} - \frac{k}{8\pi}\Gamma^{(e)\,\alpha}W_{\alpha}^{(e)}\right],
\end{align}
and the scalar master partition function is 
\begin{align}
    \mathcal{Z}_{\mathcal{N}=1}^{\text{Scalar}} &=\int \mathcal{D}V_{0}^{(m)}\mathcal{D}S^{(m)}\mathcal{D}V_{0}^{(e)}\mathcal{D}S^{(e)}\mathcal{D}\Lambda_{(0)}^{(S)}\mathcal{D}\Lambda_{(2)}^{(S)}\exp\left( i \, S_{\mathcal{N}=1}^{\text{Scalar}}\right),
\end{align}
where
\begin{align}
    S_{\mathcal{N}=1}^{\text{Scalar}}=\int \mathrm{d}^{2}\theta\int \mathrm{d}^{3}x\,&\left[\frac{2g^{2}}{\left(2\pi\right)^{2}}\left(V_{0}^{(m)}-\Lambda_{0}^{(S)}\right)\left(S^{(m)}-\Lambda_{(2)}^{(S)}\right) \right.\nonumber\\&\left.+ \frac{1}{\pi}\left(S^{(m)}-\Lambda_{(2)}^{(S)}\right)\Sigma_{(e)} + \frac{k}{2\pi}S^{(e)}\Sigma^{(e)}\right].
\end{align}
For non-zero $k$, the $\mathcal{N}=2$ magnetic change of variables \eqref{Eq: Supersymmetric change of variables} induces a shift in all three $\mathcal{N}=1$ components: 
\begin{equation}
    \left(V_{(0)}^{(e)},\,\Gamma_{\alpha}^{(e)},\,S^{(e)}\right)\rightarrow \left(V_{(0)}^{(B)},\,\Gamma_{\alpha}^{(B)},\,S^{(B)}\right) - \frac{1}{k}\left(V_{(0)}^{(m)}-\Lambda_{(0)}^{(S)},\,\Gamma_{\alpha}^{(m)}-\Lambda_{(1)\,\alpha}^{(S)},\,S^{(m)}-\Lambda_{(2)}^{(S)}\right).
\end{equation}
This yields the partition functions
\begin{align}
    \mathcal{Z}_{\text{Magnetic},\,\mathcal{N}=1}^{k\neq 0,\,\text{Vector}}&= \int \mathcal{D}\Gamma^{(m)}
    \mathcal{D}\Lambda_{(1)}^{(S)}\,\mathcal{D}\Gamma^{(B)}\,\exp i \int \mathrm{d}^{2}\theta\,\,\int\,\mathrm{d}^{3}x\,  \left[-\frac{g^{2}}{2\left(2\pi\right)^{2}}\left(\Gamma^{(m)}-\Lambda_{(1)}^{(S)}\right)^{2}\right.\nonumber\\&\qquad\qquad\qquad\qquad\left. + \frac{1}{8\pi k }\left(\Gamma^{(m)\,\alpha}-\Lambda_{(1)}^{(S)\,\alpha}\right)W^{(m)\,\alpha} - \frac{ k}{8\pi}\Gamma^{(B)\,\alpha}W^{(B)}_{\alpha}\right],
\end{align}
and
\begin{align}
    \mathcal{Z}_{\text{Magnetic},\,\mathcal{N}=1}^{k\neq 0,\;\text{Scalar}}
    &= \int 
        \mathcal{D}V_{0}^{(m)}\,\mathcal{D}\Lambda_{(0)}^{(S)}\,
        \mathcal{D}S^{(m)}\,\mathcal{D}\Lambda_{(2)}^{(S)}\,
        \mathcal{D}V_{0}^{(B)}\,
        \mathcal{D}S^{(B)}\,
        \nonumber\\&\exp\left\{
            i\int\! \mathrm{d}^{2}\theta\,\mathrm{d}^{3}x\;
            \Big[
                \frac{2g^{2}}{(2\pi)^{2}}\left(V_{0}^{(m)}-\Lambda_{(0)}^{(S)}\right)\left(S^{(m)}-\Lambda_{(2)}^{(S)}\right)\right. \nonumber \\&\left.+\frac{k}{2\pi}\left(S^{(B)}+\frac{1}{k}\left(S^{(m)}-\Lambda_{(2)}^{(S)}\right)\right)\left(\Sigma^{(B)} - \frac{1}{k}\left(\Sigma^{(m)} - \Sigma^{(S)}\right)\right)
            \Big]
        \right\},
\end{align}
with
\begin{equation}
    \Sigma^{(S)} = \frac{1}{2}D^{\alpha}\Lambda_{(1)\,\alpha}^{(S)}.
\end{equation}
For non-zero $g^{2}k$, both the vector and scalar branches realise an $\mathcal{N}=1$ massive dual structure. In the vector block, the electric description arises by integrating out $\Gamma^{\alpha\,(m)}-\Lambda_{(1)}^{(S)\,\alpha}$ yielding a Maxwell--Chern--Simons action that propagates a single $\mathcal{N}=1$ vector multiplet of mass $M = \frac{g^{2}k}{2\pi}$. On the magnetic side, the Stückelbergised magnetic vector produces the same topologically massive Deser--Jackiw--Chern--Simons structure as in the $\mathcal{N}=2$ analysis, now reduced to an $\mathcal{N}=1$ vector multiplet. The propagating Deser--Jackiw dynamics come from the Stückelberg-modified vector multiplet, while the decoupled level-$k$ Chern--Simons multiplet carries the gauge and topological data. Thus both electric and magnetic descriptions agree: both contain the same massive $\mathcal{N}=1$ vector multiplet with its matching Chern--Simons topological sector. Meanwhile, in the scalar block, a massive $\mathcal{N}=1$ real scalar of the same mass appears. In the electric description this multiplet is the scalar component of the Maxwell--Chern--Simons theory, whereas in the magnetic description it is realised by the Stückelberg pair $\left(V_{0}^{(m)}-\Lambda_{(0)}^{(S)}\right)\left(S^{(m)}-\Lambda_{(2)}^{(S)}\right)$, together with the scalar auxiliary contribution of the $\mathcal{N}=1$ Chern--Simons multiplet. In both cases, the propagating content reduces to a single massive real scalar multiplet of mass $M$, in agreement with the vector branch.

At the strict point $g^{2} = 0$, both the vector and scalar branches collapse to a purely topological $\mathcal{N}=1$ dual structure, with the only remaining non-trivial structure residing in the global gauge sector of the vector block. In the vector block, the electric description arises upon integrating out $\Gamma^{\alpha\,(m)}-\Lambda_{(1)}^{(S)\,\alpha}$, which imposes the flatness condition $W^{(e)}_{\alpha}=0$, yielding a constrained $\mathcal{N}=1$ level-$k$ Chern--Simons path integral; no propagating vector degrees of freedom survive. On the magnetic side, the Stückelberg degrees of freedom decouple at $g^{2} = 0$, and the remaining vector fields give the $\mathcal{N}=1$ Gaiotto-Witten-like representation of the same constrained endpoint. Both descriptions therefore match as equivalent electric and magnetic representations of the same $\mathcal{N}=1$ topological endpoint, with no local dynamics but the same topological content. In the scalar block, the electric description is the non-propagating scalar multiplet of the constrained $\mathcal{N}=2$ Chern--Simons theory, while in the magnetic description the same auxiliary structure appears, now entering through the Gaiotto--Witten interaction. In both descriptions no propagating scalar degrees of freedom remain: the entire scalar branch collapses to its constrained Chern--Simons auxiliary multiplet, fully consistent with the topological character of the $g^{2} =0$ point, where all local dynamics are absent. 

For $k=0$, both the vector and scalar branches become fully dynamical on both sides of the duality, with the entire non-topological content residing in the Maxwell sector. In the vector block, on the electric side, the absence of the Chern--Simons term leaves a pure $\mathcal{N}=1$ Maxwell action, which propagates a single massless vector multiplet. On the magnetic side, integrating out $\Gamma_{\alpha}^{(e)}$ imposes a flatness condition on the magnetic field strength, allowing the magnetic vector to be written locally as the derivative of a dual scalar. The Stückelberg multiplet realises this dual photon together with its fermionic partner, reconstructing the same $\mathcal{N}=1$ vector degrees of freedom as on the electric side. Both descriptions therefore coincide: the vector sector carries the full dynamical content of the $k=0$ theory, and with the Chern--Simons term absent, no topological contributions remain. In the scalar block, the electric side is simply the real scalar multiplet of the $\mathcal{N}=2$ Maxwell theory, while on the magnetic side it is realised by the compact scalar in the Stückelberg multiplet, together with its fermionic partner. In both descriptions, the propagating content consists of a single massless real scalar multiplet: on the magnetic side this scalar plays the role of the $\mathcal{N}=1$ supersymmetric dual photon - its Stückelberg origin ensuring the correct compactness - whereas on the electric side it appears as the ordinary Maxwell scalar. Both theories are gauge invariant and contain no topological terms in the scalar sector. Thus the scalar branch carries the full local dynamical content of the $k=0$ theory, paralleling the vector branch, where the Maxwell field likewise remains dynamical and devoid of topological terms.

\subsection{Witten index}

A consistency check of the proposed duality, a necessary condition, is that the Witten index is identical in both the electric and magnetic theory. We first consider the massive case, $g^2 \ne 0, k\ne 0$. The Witten index cannot change by a continuous variation of the coupling that preserves the mass gap. 

The Witten index of the electric theory, Maxwell--Chern--Simons, is $I_w=|k|$ 
\cite{Witten:1999ds}. It is given by the Witten index of $\mathcal{N}=2$, level $k$ Abelian super Chern--Simons theory, because the mass of the photon, generated by the gauge coupling, is irrelevant. 

On the magnetic side, the calculation is carried out while keeping $g^{2}\neq 0$. The magnetic theory consists of two decoupled sectors. A massive vector multiplet, irrelevant for the infra-red dynamics, and a pure super Chern-Simons sector, which is identical to the electric side. Thus the Witten index of the magnetic theory is also $I_w=|k|$.

The brane realisation of the two theories provides an independent check of the finite-$g^{2}$ result $I_{w}=|k|$. In both cases it is given by the combinatorics of choosing a single fivebrane out of a collection of $k$ fivebranes \cite{Ohta:1999iv}.

Since both $V^{(e)}$ and $V^{(m)}$ are dynamical compact $\mathrm{U}(1)$ vector multiplets, at the strict point $g^{2}=0$ the magnetic endpoint is the $\mathrm{U}(1)^{2}$ BF--Chern--Simons theory with $K$-matrix
\begin{equation}
    K=
    \begin{pmatrix}
        0 & 1\\
        1 & k
    \end{pmatrix}.
\end{equation}
Canonical quantisation on $T^{2}$ gives $I_{w} = |\det K| = 1$\footnote{We thank the anonymous referee for pointing it out.}. On the electric side, integration over $V^{(m)}$ gives the constrained Chern--Simons representation of the same strict endpoint by imposing $W^{(e)}=0$. Thus the Witten index is $I_{w}=1$ on both sides at $g^{2}=0$. 

The limit $g^{2}\to 0$ is nevertheless subtle. For every finite $g^{2}>0$ with $k\neq 0$, the dual pair flows in the IR to level-$k$ Chern--Simons theory and has $I_{w}=|k|$. As $g^{2}\to 0$, the mass $M = \frac{g^{2}k}{2\pi}$ tends to zero and the mass gap closes, so the Witten index is no longer protected by continuous variation of the coupling. The strict $g^{2}=0$ endpoint must therefore be distinguished from the dual pair at finite $g^{2}>0$. There is no contradiction between $I_{w}=|k|$ for the finite-$g^{2}$ dual pair and $I_{w}=1$ at the strict endpoint; equivalently, the infrared limit and the $g^{2}\to0$ limit need not commute.

Note also that the $k=0$ theory is a gapless free theory of vector multiplet dual to a chiral multiplet, and therefore lies outside the massive Witten index discussion above.

\subsection{Global structure and lessons for the non-Abelian case}
\subsubsection*{Global structure and line-operator matching}
Both the electric and magnetic theories admit Wilson and 't Hooft line operators, and these are objects sensitive to the global form of the gauge group. Two gauge theories may share the same Lie algebra but differ in global structure. For example $G$ versus $G/\mathbb{Z}_{k}$, or $\mathrm{SU}(2)$ versus $\mathrm{SO}(3) = \mathrm{SU}(2)/\mathbb{Z}_{2}$. Although these theories are locally indistinguishable, they differ physically in the line operators they permit; an $\mathrm{SO}(3)$ gauge theory cannot admit a fundamental Wilson line, whereas an $\mathrm{SU}(2)$ gauge theory can. Moreover, such line operators acquire non-trivial phases under large gauge transformations, and it is these phases that encode the global (rather than Lie-algebraic) information.

A duality between an electric theory with gauge group $G_{e}$ and a magnetic theory with gauge group $G_{m}$ does not require $G_{e} = G_{m}$. Instead, the duality requires
\begin{enumerate}
    \item \textbf{Matching spectra of line operators}\newline
    The Wilson and 't Hooft lines permitted by the global form of $G_{e}$ must map to the lines permitted by the global form of $G_{m}$. Two theories with different global forms may still be dual if the sets of allowed line operators, together with their charges, match.
    \item \textbf{Matching large-gauge transformation phases}\newline
    Line operators acquire phases under large gauge transformations determined by the topology of the gauge fields. These phases must be preserved by the duality map.
    \item \textbf{Interchange of charges}\newline
    Electric and magnetic global symmetries are exchanged under the duality.  
    In the Abelian theory, the electric 1-form current $F$ is exchanged with the magnetic 0-form current $\star F$. Accordingly, Wilson and 't Hooft line operators exchange roles as objects charged under these respective symmetries.
\end{enumerate}
The Abelian construction does not literally enforce $G_{e} = G_{m}$; for Chern--Simons level $k=0$ the magnetic side is not a gauge theory at all. Instead, the Stückelberg mechanism makes the dual scalar compact $\sigma\sim\sigma+2\pi$, so that its values lie in a $\mathrm{U}(1)$ target. This ensures that the magnetic description carries the same $\mathrm{U}(1)$ holonomy data as the electric gauge theory, even though only the electric side has a genuine $\mathrm{U}(1)$ gauge symmetry.

In this sense, the Abelian Stückelberg mechanism aligns the global structure of the electric and magnetic descriptions by ensuring that their holonomy sectors match, even though only one side is a genuine gauge theory. In the non-Abelian parent theory introduced later, we will choose a single gauge symmetry acting on all multiplets; this has the effect of enforcing $G_{e} = G_{m}$. Allowing $G_{e}\neq G_{m}$ in the non-Abelian case is in principle possible, but implementing this would require additional structure that we do not develop here. Our construction is the simplest controlled non-Abelian extension. This is a sufficient and convenient choice that guarantees matching line-operator spectra and large-gauge phases automatically but is stronger than necessary: non-Abelian dualities may relate distinct global forms of the same Lie algebra provided their line-operator data match. Thus, the Stückelberg approach provides a sufficient, though not strictly necessary, mechanism for ensuring global structure compatibility across the duality.

\subsubsection*{Why compactness and group-valued Stückelberg fields are essential}
It is useful to explain why group valued Stückelberg fields are required to encode the correct global data. 

In the Abelian theory, the correct statement is that Maxwell--Chern--Simons is dual not to the gauge Deser--Jackiw model, but to a non-gauge Deser--Jackiw vector plus a decoupled Chern--Simons sector. This separation restores the correct topological behaviour and the correct action of large gauge transformations. In our Abelian $\mathcal{N}=2$ construction, this logic is built in from the start.

In the Abelian case with $k=0$, integrating out the electric gauge field sets $F^{(m)}=0$, and so locally $A^{(m)} = \mathrm{d}\sigma$. If one takes $\sigma\in \mathbb{R}$, this reproduces only the local equations. To recover the correct global $\mathrm{U}(1)$ structure, $\sigma$ must be compact, $\sigma\sim\sigma+2\pi$, or equivalently that $e^{i\sigma}$ be single-valued. Flat $\mathrm{U}(1)$ connections are classified by holonomies 
\begin{equation}
    \mathrm{Hol}\left(\gamma\right) = \exp\left(i\oint_{\gamma}A\right)\in \mathrm{U}(1),
\end{equation}
which define a homomorphism $\rho: \pi_{1}\left(\mathcal{M}\right)\to \mathrm{U}(1)$, well defined up to gauge equivalence. On manifolds without torsion, the moduli space of flat $\mathrm{U}(1)$ connections is given by \cite{hatcher2005algebraic}
\begin{equation}
    \label{Eq: Abelian isomorphism}
    \mathrm{Hom}\left(\pi_{1}\left(\mathcal{M}\right),\,\mathrm{U}(1)\right)\cong H^{1}\left(\mathcal{M};\,\mathrm{U}(1)\right)\cong H^{1}\left(\mathcal{M};\,\mathbb{R}\right)/2\pi H^{1}\left(\mathcal{M},\,\mathbb{Z}\right),
\end{equation}
that is, every flat $\mathrm{U}(1)$ connection is completely characterised by its holonomies around closed loops. A single compact scalar $\sigma\in \mathbb{R}/2\pi\mathbb{Z}$ therefore captures the full holonomy moduli space. The Stückelberg formulation enforces this compactness automatically, guaranteeing that the magnetic theory carries the full global structure of the electric one.

In the non-Abelian case, no analogue of the Abelian isomorphism \eqref{Eq: Abelian isomorphism} exists; we cannot in general recover group level data from the Lie-algebra. Flat $G$-connections are identified by group-valued holonomies
\begin{equation}
    \mathrm{Hol}\left(\gamma\right) = \mathcal{P}\exp\left(\oint_{\gamma}A\right)\in G,
\end{equation}
which are genuinely group valued. Because these maps define a homomorphism $\rho: \pi_{1}\left(\mathcal{M}\right)\to G$, well defined up to conjugation; flat $G$-connections are classified by $\mathrm{Hom}\left(\pi_{1}\left(\mathcal{M}\right),\,G\right)/G$.  Because the exponential map $\exp:\mathfrak{g}\to G$ fails to be globally invertible, and because path ordering is essential, a Lie-algebra valued description can never encode all flat sectors or their large-gauge-transformation behaviour.

This motivates the use of a genuinely group-valued Stückelberg multiplet in the non-Abelian parent theory: it guarantees that when we impose the flatness constraint in the parent action, the magnetic description automatically inherits the full non-Abelian holonomy data required for duality.

%% file: sections/4_nonAbelian.tex
\section{Non-Abelian S-duality}
\label{Section: Non-Abelian S-duality}
\subsection{\texorpdfstring{$\mathcal{N}=2$}{N=2} supersymmetry}
We non-Abelianise the $\mathcal{N}=2$ master partition function by promoting all fields to Lie-algebra-valued superfields. We first promote the two vector superfields $V^{(m)}$ and $V^{(e)}$ to be Lie algebra valued fields via
\begin{equation}
    V^{(m)} = V_{(m)}^{A}T^{A},\quad V^{(e)}=V_{(e)}^{A}T^{A},
\end{equation}
where $T^{A}$ are the Hermitian generators of $\mathfrak{u}\left(N\right)$, 
\begin{equation}
    \left[T^{A},\,T^{B}\right] = i f^{ABC}T^{C},
\end{equation}
so that $V^{\dagger}=V$ and $e^{V}\in \mathrm{U}\left(N\right)$. The structure constants are normalised by
\begin{equation}
    \mathrm{Tr}\left(T^{A}T^{B}\right) =\frac{1}{2}\delta^{AB}.
\end{equation}
We likewise promote the chiral fields 
\begin{equation}
    \Omega = \Omega^{A}T^{A},
\end{equation}
and continue to define $\Lambda^{(S)} = i\left(\Omega^{(S)}-\Omega^{\dagger\,(S)}\right)$. We upgrade the Abelian field strength $W(V) = \frac{i}{2}\bar{D}DV$ to the non-Abelian
\begin{equation}
    W(V)= \frac{i}{2}\int_{0}^{1}\,\mathrm{d}s\bar{D}^{\alpha}\left(e^{-sV}\left(D_{\alpha}V\right)e^{sV}\right)=\frac{i}{2}\,\bar D^\alpha\!\Big(e^{-V}D_\alpha e^{V}\Big).
\end{equation}
To justify the second equality here, we introduce a superfield $f$ dependent on a parameter $s$ through $f(s) = e^{-sV}\left(D_{\alpha}e^{sV}\right)$. Differentiating with respect to $s\in \left[0,\,1\right]$ and using the product rule gives
\begin{equation}
    \frac{d}{ds}f(s) = -e^{-sV}VD_{\alpha}e^{sV} + e^{-sV}D_{\alpha}\left(Ve^{sV}\right) = e^{-sV}\left(D_{\alpha}V\right)e^{sV}.
\end{equation}
Integrating from $s=0$ to $s=1$ therefore yields
\begin{equation}
    \int_{0}^{1}e^{-sV}\left(D_{\alpha}V\right)e^{sV}= f(1) - f(0) = e^{-V}D_{\alpha}e^{V} - e^{0}D_{\alpha}e^{0} = e^{-V}D_{\alpha}e^{V}.
\end{equation}
Substituting this into the superspace definition of the non-Abelian field strength then gives the expression $W(V) = \frac{i}{2}\bar{D}^{\alpha}\left(e^{-V}D_{\alpha}e^{V}\right)$. Gauge transformations are implemented on the fields via a group element $g$, which we write
\begin{equation}
    g = e^{i\Lambda_{g}},
\end{equation}
with $\Lambda_{g}$ chiral; $\bar{D}^{\alpha}\Lambda_{g}=0$. The vector multiplets are transformed by the rule
\begin{equation}
    e^{V}\longrightarrow g^{\dagger}e^{V}g,
\end{equation}
and the Stückelberg fields are transformed via
\begin{equation}
    e^{\Lambda}\longrightarrow g^{\dagger} e^{\Lambda}g.
\end{equation}
It is straightforward to verify that under a gauge transformation $e^{V}\to g^{\dagger}e^{V}g$ with $g$ chiral ($\bar D_\alpha g=0$), that
\begin{equation}
    e^{-V}D_{\alpha}e^{V}\rightarrow g^{\dagger}\big(e^{-V}D_\alpha e^{V}\big)g \;+\; g^{\dagger}D_\alpha g. 
\end{equation}
Since the second term is chiral, its $\bar{D}^{\alpha}$ derivative vanishes and so $W\left(V\right)$ is gauge covariant,
\begin{equation}
    W\left(V\right) = \frac{i}{2}\bar{D}^{\alpha}\left(e^{-V}D_{\alpha}e^{V}\right)\longrightarrow g^{\dagger}W(V)g.
\end{equation}
The non-Abelian field strength correctly reduces to the Abelian field strength $W = \frac{i}{2}\bar{D}DV$ in the Abelian limit, ensuring that the non-Abelian construction is a direct supersymmetric extension of the Abelian duality framework.

Having established the underlying non-Abelian supersymmetric multiplets and field strengths, we generalise the Abelian $V_{(m)}'$ via
\begin{equation}
    V'^{(m)} = \int_{0}^{1}\mathrm{d}s\,U^{-1}\left(\partial_{s}U\right),\quad U(s)=e^{-s \Lambda^{(S)}}e^{s V^{(m)}},\quad s\in \left[0,\,1\right],
\end{equation}
which under a gauge transformation transforms covariantly as
\begin{equation}
    V'^{(m)}\rightarrow g^{\dagger}V'^{(m)}g.
\end{equation}
In the Abelian limit $V'^{(m)}$ correctly reproduces $V^{(m)}-\Lambda^{(S)}$.

From here, we non-Abelianise each of the three terms in the master partition function. Crucially, the master action contains the mixed term $V'^{(m)}W^{(e)}$. This term is gauge covariant only if both vector multiplets transform under the same non-Abelian gauge transformation. Thus, the parent theory carries a single non-Abelian gauge symmetry acting on $V'^{(m)}$ and $V^{(e)}$. With this understood, the non-Abelian parent of the master partition function reads
\begin{align}
    \mathcal{\hat{Z}}= \int \mathcal{D}V_{(m)}\mathcal{D}V_{(e)}\mathcal{D}\Lambda_{(S)}\exp i \int \mathrm{d}^{4}\theta\,\int\,\mathrm{d}^{3}x\,\mathrm{Tr} &\left[\frac{2g^{2}}{\left(2\pi\right)^{2}}V_{(m)}^{'2}+\frac{2
    i}{\pi}\,V'^{(m)}W^{(e)}+\,\frac{ik}{\pi}V^{(e)}W^{(e)}\right].
\end{align}
Since $W(V)$ reduces to the Abelian field strength in the limit $\mathfrak{u}\left(N\right)\rightarrow\mathfrak{u}\left(1\right)$, the non-Abelian parent partition function collapses to the Abelian master partition function as required. 

Our non-Abelian Stückelberg construction is designed precisely to accommodate the global data that cannot be captured at the Lie-algebra level. The magnetic superfield $V'^{(m)}$ is adjoint-valued and transforms covariantly under the single non-Abelian gauge symmetry of the parent theory, but it is deliberately not a gauge connection. The Stückelberg multiplet is taken to be genuinely group-valued, and it is this choice that guarantees that the magnetic variables inherit the correct global topological behaviour. 

Consequently, when the flatness constraint is imposed in the master action, the remaining degrees of freedom span the full moduli space $\mathrm{Hom}\left(\pi_{1}\left(\mathcal{M}\right),\,G\right)/G$ of flat $G$-connections, with the correct conjugation action. A purely Lie-algebra formulation would miss this information. This is the non-Abelian analogue of the Abelian fact that the Stückelberg implementation of the dual photon automatically builds in compactness and therefore the full $\mathrm{U}(1)$ holonomy data.

\subsubsection{Electric side behaviour}
For the $g^{2}\neq 0 $ case, we integrate over $V_{m}'$ to obtain
\begin{align}
    \mathcal{\hat{Z}}_{\text{Electric}}^{g^{2}\neq 0 }= \int\mathcal{D}V^{(e)}\exp i \int \mathrm{d}^{4}\theta\,\int\,\mathrm{d}^{3}x\,\mathrm{Tr} &\left[\frac{2}{g^{2}}W_{(e)}^{2}+i\,\frac{k}{\pi}V^{(e)}W^{(e)}\right],
\end{align}
where the field-independent Gaussian determinant has been absorbed into the normalisation of $\mathcal{\hat{Z}}$. This is $\mathcal{N}=2$ Yang-Mills--Chern-Simons theory. This reduces to pure $\mathcal{N}=2$ Yang-Mills for the special case $k=0$ as
\begin{align}
    \mathcal{\hat{Z}}_{\text{Electric}}^{k=0}= \int \mathcal{D}V^{(e)}\exp i \int \mathrm{d}^{4}\theta\,\int\,\mathrm{d}^{3}x\,\mathrm{Tr} &\left[\frac{2}{g^{2}}W_{(e)}^{2}\right].
\end{align}
In the case of $g^{2}=0$, the Gaussian term in the master action is removed and so integration over $V'^{(m)}$ imposes the flatness constraint
\begin{equation}
    \int\mathcal{D}V'^{(m)}\exp i \int \mathrm{d}^{4}\theta\,\int\mathrm{d}^{3}x\,\mathrm{Tr}\left(\frac{2i}{\pi}V'^{(m)}W^{(e)}\right)\propto \delta\left(W^{(e)}\right),
\end{equation}
leaving a constrained non-Abelian Chern-Simons path integral
\begin{align}
    \mathcal{\hat{Z}}= \int \mathcal{D}V^{(e)}\delta\left(W^{(e)}\right)\exp i \int \mathrm{d}^{4}\theta\,\int\,\mathrm{d}^{3}x\,\mathrm{Tr} &\left[\,\frac{ik}{\pi}V^{(e)}W^{(e)}\right].
\end{align}
As in the Abelian case, the Dirac delta enforces flatness, but allows nontrivial holonomy sectors.

\subsubsection{Magnetic side behaviour}
\subsubsection*{Non-zero $k$}
As in the Abelian case, for non-zero $k$, we may substitute the usual
\begin{equation}
\label{Eq:Usual-change-of-variables}
V^{(e)} \;=\; B \;-\; \frac{1}{k}\,V'^{(m)},
\end{equation}
to obtain the magnetic side partition function. Because the field strength $W(V)$ depends nonlinearly on the underlying field $V$ through
\begin{equation}
    W\left(V\right) = \frac{i}{2}\bar{D}^{\alpha}\Gamma_{\alpha}\left(V\right),\quad \Gamma_{\alpha}\left(V\right) = e^{-V}D_{\alpha}e^{V},
\end{equation}
we separate the Abelian-like and non-Abelian interaction contributions by defining
\begin{equation}
    I\left(X,\,Y\right) = W\left(X+Y\right) - W\left(X\right) - W\left(Y\right).
\end{equation}
Observe that $I(X,Y) = I(Y,X)$. In the Abelian limit, $W_{\mathrm{U(1)}}\left(X+Y\right) = W_{\mathrm{U(1)}}\left(X\right)+W_{\mathrm{U(1)}}\left(Y\right)$ and so $I\left(X,\,Y\right)$ vanishes. 

Since all non-Abelian corrections come from commutators, we introduce the nested adjoint action
\begin{equation}
     [V,\cdot]^0 X := X,\qquad
    [V,\cdot]^{n+1} X := [V,\,[V,\cdot]^n X],
\end{equation}
such that 
\begin{equation}
     \left[V,\,\cdot\right]^{n} X :=
 \underbrace{[V,[V,\dots,[V, X]\dots]]}_{n\ \text{commutators}}.
\end{equation}
We then understand exponentials of commutators as a formal power series
\begin{equation}
    e^{\left[X,\,\cdot\right]} = \sum_{n=0}^{\infty}\frac{\left[X,\,\cdot\,\right]^{n}}{n!}.
\end{equation}
With this, to derive $I\left(X,\,Y\right)$, we define
\begin{equation}
    \Phi_{Z}\left(\cdot\right) = \int_{0}^{1}\mathrm{d}s\,e^{-s\left[Z,\,\cdot\,\right]} = \frac{1-e^{-\left[Z,\,\cdot\,\right]}}{\left[Z,\,\cdot\,\right]}. 
\end{equation}
We then have that
\begin{equation}
    \Gamma_{\alpha}\left(X+Y\right)- \Gamma_{\alpha}\left(X\right)- \Gamma_{\alpha}\left(Y\right) = \left(\Phi_{X+Y}-\Phi_{X}\right)\left(D_{\alpha}X\right) +    \left(\Phi_{X+Y}-\Phi_{Y}\right)\left(D_{\alpha}Y\right).
\end{equation}
To evaluate this, we then apply the Duhamel formula
\begin{equation}
    e^{-s(A+B)}-e^{-sA}=-\int_0^s d\tau\,e^{-(s-\tau)A}\,B\,e^{-\tau(A+B)}
\end{equation}
to the exponential $e^{-s\left[X+Y,\,\cdot\,\right]}$ inside $\Phi_{X+Y}$. We obtain
\begin{align}
    \Phi_{X+Y}-\Phi_{X}
    &=-\int_{0}^{1}\mathrm{d}s\,\int_{0}^{s}\mathrm{d}\tau\,e^{-\left(s-\tau\right)\left[X,\,\cdot\right]}\left[Y,\,\cdot\,\right]\,e^{-\tau\left[X+Y,\,\cdot\,\right]}.
\end{align}
Finally then,
\begin{equation}
    I\left(X,\,Y\right) = \frac{i}{2}\bar{D}^{\alpha}\left(\left(\Phi_{X+Y}-\Phi_{X}\right)D_{\alpha}X +    \left(\Phi_{X+Y}-\Phi_{Y}\right)D_{\alpha}Y\right).
\end{equation}
We note that $I\left(X,\,Y\right)$ vanishes whenever either $X$ or $Y$ is central. If $X$ and $Y$ transform covariantly under the same gauge transformation, then $I\left(X,\,Y\right)\rightarrow g^{\dagger}I\left(X,\,Y\right)g$ also transforms covariantly. Finally, $I\left(X,\,Y\right)$ admits a series expansion in nested colour commutators, given in appendix \ref{Appendix:Integrating-the-field-strength-interactions}.

Specialising this to $V^{(e)}=B-\dfrac{1}{k}V'^{(m)}$, we let $X=B$, $Y=-\dfrac{1}{k}V'^{(m)}$. We split
\begin{equation}
  W\!\left(B-\tfrac{1}{k}V'^{(m)}\right)
  = W(B)\;+\;W\!\left(-\tfrac{1}{k}V'^{(m)}\right)\;+\;I\!\left(B,-\tfrac{1}{k}V'^{(m)}\right).
\end{equation}
This decomposition isolates the nonlinear gauge-covariant interaction encoded in $I$. This leaves
\begin{align}
    S
    &= \int d^{4}\theta\int d^{3}x\,\mathrm{Tr}\Bigg[
    \frac{2g^{2}}{(2\pi)^{2}}\,V_{(m)}^{'2}
    +\frac{i}{\pi}\,V'^{(m)}\,
     W\,\Big(-\tfrac{1}{k}V'^{(m)}\Big)
    +\frac{ik}{\pi}\,B\,W(B)\nonumber\\
    &+\;\frac{i}{\pi}\,\Big(V'^{(m)}+kB\Big)\,
   I\!\Big(B,-\tfrac{1}{k}V'^{(m)}\Big)+\;\frac{i}{\pi}\,\Big(\,V'^{(m)}\,W(B)
    \;+\;k\,B\,W\!\Big(-\tfrac{1}{k}V'^{(m)}\Big)\Big)
    \Bigg].
\end{align}
In the Abelian limit, the interaction term $I$ vanishes, correctly reducing this to the Abelian master action. In this limit, the cross terms linear in $W\left(B\right)$ and $W\left(-\frac{1}{k}V'^{(m)}\right)$ combine into a total superspace derivative and can be dropped, as in the bosonic non-Abelian case. Away from the Abelian limit, this combination is not a total derivative and remains dynamical. Thus the partition function takes the form
\begin{align}
    \mathcal{\hat{Z}}= \int \mathcal{D}V^{(m)}\mathcal{D}V^{(e)}\mathcal{D}\Lambda^{(S)}&\exp i \int \mathrm{d}^{4}\theta\,\int\,\mathrm{d}^{3}x\,\mathrm{Tr} \left[\frac{2g^{2}}{\left(2\pi\right)^{2}}V_{(m)}^{'2}+ \frac{i}{\pi}V'^{(m)}W\left(-\frac{1}{k}V'^{(m)}\right)\right.\nonumber\\&\left.+ \frac{ik}{\pi}BW\left(B\right)+ \frac{i}{\pi}\left(V'^{(m)}\left(I\left(B,\,-\frac{1}{k}V'^{(m)}\right)+W(B)\right)\right.\right.\nonumber\\
    &\left.\left.+kB\left(I\left(B,\,- \frac{1}{k}V'^{(m)}\right) + W\left(-\frac{1}{k}V'^{(m)}\right)\right)\right)\right].
\end{align}
This furnishes the non-Abelian supersymmetric extension of the Deser--Jackiw master partition function, reducing correctly to the Abelian case and revealing an inherently non-Abelian interaction sector encoded by $I\left(X,\,Y\right)$.

\subsubsection*{$g^{2}=0$ case}
In the case of $g^{2}=0$, the quadratic term is removed from the partition function, leaving
\begin{align}
    \mathcal{Z}= \int \mathcal{D}V^{(m)}\mathcal{D}V^{(e)}\mathcal{D}\Lambda^{(S)}\exp i \int \mathrm{d}^{4}\theta\,\int\,\mathrm{d}^{3}x\, &\mathrm{Tr}\left[\frac{2
    i}{\pi}\,V'^{(m)}W^{(e)}+\frac{ik}{\pi}V^{(e)}W^{(e)}\right].
\end{align}
Since $V'^{(m)}$ appears linearly, integrating it out imposes $\delta\left(W\left(V^{(e)}\right)\right)$ which constrains the electric field strength to be flat. To make this manifest, we remove the Stückelberg degree of freedom by a change of variables: we act on $V^{(e)}$ and $V'^{(m)}$ by a chiral group element $g = e^{\Omega}$, with $\bar{D}_{\alpha}\Omega=0$, as a gauge transformation would, but we do not transform $\Lambda^{(S)}$ (so this is a field redefinition rather than a gauge transformation).
\begin{align}
    & e^{\tilde V^{(e)}}:= g^{\dagger}\,e^{V^{(e)}}\,g
    \quad\Rightarrow\quad
    W(\tilde V^{(e)})=g^{\dagger}W(V^{(e)})g,\\
    & \tilde V^{(m)}:=g^{\dagger}V'^{(m)}g
    \quad\Leftrightarrow\quad
    V'^{(m)}=g\,\tilde V^{(m)}\,g^{\dagger}.
\end{align}
Since $V'^{(m)}$ already contains $\Lambda^{(S)}$, this change of variables disentangles $V'^{(m)}$ from $\Lambda^{(S)}$, causing $\Lambda^{(S)}$ to drop out of the action. The resulting path integral over $\Lambda^{(S)}$ is trivial, leaving a proposed non-Abelian $\mathcal{N}=2$ Gaiotto--Witten-like representation of the constrained endpoint \cite{Gaiotto:2008sa}:
\begin{align}
    \mathcal{Z}= \int \mathcal{D}\tilde{V}^{(m)}\,\mathcal{D}\tilde{V}^{(e)}\exp i \int \mathrm{d}^{4}\theta\,\int\,\mathrm{d}^{3}x\, &\mathrm{Tr}\left[\frac{2
    i}{\pi}\,\tilde{V}^{(m)}W\left(\tilde{V}^{(e)}\right)+\frac{ik}{\pi}\tilde{V}^{(e)}W\left(\tilde{V}^{(e)}\right)\right].
\end{align}
This provides a magnetic representation of the same constrained supersymmetric Chern--Simons endpoint, with holonomy sectors enforced by the flatness constraint.

\subsubsection*{$k=0$ case}
In the case of $k=0$, the master partition function reduces to
\begin{align}
    \label{Eq: k=0 partition function}
    \mathcal{\hat{Z}}^{k=0}= \int \mathcal{D}V_{(m)}\mathcal{D}V_{(e)}\mathcal{D}\Lambda_{(S)}\exp i \int \mathrm{d}^{4}\theta\,\int\,\mathrm{d}^{3}x\,\mathrm{Tr} &\left[\frac{2g^{2}}{\left(2\pi\right)^{2}}V_{(m)}^{'2}+\frac{2
    i}{\pi}\,V'^{(m)}W^{(e)}\right].
\end{align}
Integration over $V^{(e)}$ is non-trivial due to the nonlinear nature of $W\left(V\right)$. Instead, it is linear in the pre-potential
\begin{equation}
    \Gamma_{\alpha}^{(e)} = e^{-V^{(e)}}D_{\alpha}e^{V^{(e)}},
\end{equation}
as
\begin{align}
    \label{Eq: Gamma linearity}
    \int \mathrm{d}^{4}\theta\,\int\mathrm{d}^{3}x\,\mathrm{Tr}\left(V'^{(m)}W^{(e)}\right) 
    &= - \frac{i}{2}\int \mathrm{d}^{4}\theta\,\int\mathrm{d}^{3}x\,\mathrm{Tr}\left(\left(\bar{D}^{\alpha}V'^{(m)}\right)\Gamma^{(e)}_{\alpha}\right)
\end{align}
up to a total superspace derivative. Starting with \eqref{Eq: k=0 partition function} and using \eqref{Eq: Gamma linearity}, we have
\begin{align}
    \mathcal{\hat{Z}}^{k=0}= \int \mathcal{D}V^{(m)}\mathcal{D}V^{(e)}\mathcal{D}\Lambda^{(S)}\exp i \int \mathrm{d}^{4}\theta\,\int\,\mathrm{d}^{3}x\,\mathrm{Tr} &\left[\frac{2g^{2}}{\left(2\pi\right)^{2}}V_{(m)}^{'2}+\frac{1}{\pi}\,\left(\bar{D}^{\alpha}V'^{(m)}\right)\Gamma^{(e)}_{\alpha}\right]. 
\end{align}
We may then introduce the group variable
\begin{equation}
    g = e^{V^{(e)}},
\end{equation}
such that
\begin{equation}
    \Gamma_{\alpha}^{(e)} = g^{\dagger}D_{\alpha}g.
\end{equation}
Because the path integral measure $\mathcal{D}V^{(e)}$ is invariant under gauge transformations of $V^{(e)}$, which act on $g = e^{V^{(e)}}$ as $g\to g e^{\epsilon}$, we may perform this as a change of variables in the electric sector without the value of the integral changing. Consequently,
\begin{equation}
    0 = \frac{\delta}{\delta \epsilon}\int \mathcal{D}V^{(e)}\,\exp\left(\frac{i}{\pi}\int \mathrm{d}^{4}\theta\int\mathrm{d}^{3}x\,\mathrm{Tr}\left(\left(\bar{D}^{\beta}V'^{(m)}\right)\Gamma_{\beta}^{(e)}\right)\right).
\end{equation}
With $\varepsilon$ chiral, we have that
\begin{equation}
    \delta\left(\Gamma_{\alpha}^{(e)}\right) = D_{\alpha}\epsilon + \left[\Gamma_{\alpha}^{(e)}, \epsilon\right].
\end{equation}
Performing the variation of the exponent, discarding total superspace derivatives, and factorising out $\epsilon$, we have
\begin{align}
    0 &= \frac{i}{\pi}\int \mathcal{D}V^{(e)}\left(\left\{\int \mathrm{d}^{4}\theta\,\int\mathrm{d}^{3}x\,\mathrm{Tr}\left(\epsilon\left[D_{\alpha}\left(\bar{D}^{\alpha}V'^{(m)}\right) + \left[\Gamma_{\alpha}^{(e)},\,\left(\bar{D}^{\alpha}V'^{(m)}\right)\right]\right]\right)\right\}\right.\nonumber\\&\qquad\qquad\times\left.\exp\frac{i}{\pi}\int\mathrm{d}^{4}\theta\int\mathrm{d}^{3}x\,\mathrm{Tr}\left(\left(\bar{D}^{\beta}V'^{(m)}\right)\Gamma_{\beta}^{(e)}\right)\right).
\end{align}
Then since $\epsilon$ is arbitrary, we must have
\begin{equation}
    \nabla^{\alpha}_{(e)}\left(\bar{D}_{\alpha}V'^{(m)}\right)  = 0,
\end{equation}
with
\begin{equation}
    \nabla^{\alpha}_{(e)}X_{\alpha} = D^{\alpha}X_{\alpha}+\left[\Gamma^{(e)\,\alpha},\,X_{\alpha}\right]\,\quad \Gamma^{(e)}_{\alpha} = e^{-V^{(e)}}D_{\alpha}e^{V^{(e)}}.
\end{equation}
In the Abelian case, this was imposed pointwise for all configurations of $V^{(e)}$ appearing in the path integral. This is allowed as in the Abelian theory $g^{\dagger}D_{\alpha}g = D_{\alpha}V^{(e)}$ and the commutator vanishes, and so the operator $\nabla^{(e)\,\alpha}\bar{D}_{\alpha}$ does not depend on $V^{(e)}$. In the non-Abelian case, demanding $\nabla^{\alpha}_{(e)}\left(\bar{D}_{\alpha}V'^{(m)}\right) = 0$ for every configuration of $V^{(e)}$ would force $\left[\Gamma^{(e)\,\alpha},\,\bar{D}_{\alpha}V^{(m)}\right]=0$ for all $\Gamma^{(e)\,\alpha}$, which in turn would force $\bar{D}_{\alpha}V'^{(m)}$ to lie in the center of the algebra. This would incorrectly collapse the dual theory to its centre. Therefore, instead of a pointwise application, we introduce 
\begin{equation}
    \delta\!\left(\mathcal \nabla^{(e)\,\alpha}\bar{D}_{\alpha}(V'^{(m)})\right)
    = \frac{1}{\det'\left(\nabla^{(e)\,\alpha}\bar{D}_{\alpha}\right)}\;
    \int \mathcal D P\;
    \exp\!\left\{\frac{i}{2\pi}\!\int d^3x\,d^4\theta\;
    \mathrm{Tr}\,\big(P\,\mathcal \nabla^{(e)\,\alpha}\bar{D}_{\alpha}(V'^{(m)})\big)\right\},
\end{equation}
where $\det'\left(\nabla^{(e)\,\alpha}\bar{D}_{\alpha}\right)$ denotes the reduced determinant that excludes states in the kernel of $\nabla^{(e)\,\alpha}\bar{D}_{\alpha}$. Gauge invariance of this representation requires the auxiliary superfield to transform in the adjoint, $P\rightarrow g^{\dagger}Pg$. This identity holds for each fixed $V^{(e)}$. In the Abelian limit, the $|\det\nabla^{(e)\,\alpha}\bar{D}_{\alpha}|$ is constant in $V^{(e)}$, which is what allowed the integration before. Consequently, we have that 
\begin{align}
    &\int \mathcal{D}V^{(e)}\,\exp\left(\frac{i}{\pi}\int\mathrm{d}^{4}\theta\int\mathrm{d}^{3}x\,\mathrm{Tr}\left(\left(\bar{D}^{\alpha}V'^{(m)}\right)\Gamma_{\alpha}^{(e)}\right)\right)\nonumber\\&\propto \frac{1}{\det'\nabla^{(e)\,\alpha}\bar{D}_{\alpha}}\;
    \int \mathcal D P\;
    \exp\!\left\{\frac{i}{2\pi}\!\int d^3x\,d^4\theta\;
    \mathrm{Tr}\,\big(P\,\mathcal \nabla^{(e)\,\alpha}\bar{D}_{\alpha}(V'^{(m)})\big)\right\}.
\end{align}
Therefore, the partition function must be
\begin{align}
    \mathcal{\hat{Z}}^{k=0}= \int \mathcal{D}V^{(m)}\mathcal{D}V^{(e)}\mathcal{D}\Lambda^{(S)}\mathcal{D}P\,\frac{1}{\det'\nabla^{(e)\,\alpha}\bar{D}_{\alpha}}\,\exp i \int \mathrm{d}^{4}\theta\,&\int\,\mathrm{d}^{3}x\,\mathrm{Tr} \left[\frac{2g^{2}}{\left(2\pi\right)^{2}}V_{(m)}^{'2}\right.\nonumber\\&\left.+\frac{i}{2\pi}P\,\mathcal \nabla^{(e)\,\alpha}\bar{D}_{\alpha}(V'^{(m)})\right]. 
\end{align}
We may expand the action around the ``Abelian part'' by writing $\nabla_{\alpha}^{(e)} = D_{\alpha} + \left[\Gamma_{\alpha}^{(e)},\,\cdot\right]$ to factorise out all terms independent of $V^{(e)}$. This yields
\begin{align}
    \mathcal{\hat{Z}}^{k=0}= &\int \mathcal{D}V^{(m)}\mathcal{D}\Lambda^{(S)}\mathcal{D}P\,\exp i \int \mathrm{d}^{4}\theta\,\int\,\mathrm{d}^{3}x\,\mathrm{Tr} \left[\frac{2g^{2}}{\left(2\pi\right)^{2}}V_{(m)}^{'2}+\frac{i}{2\pi}P\, D^{\alpha}\bar{D}_{\alpha}(V'^{(m)})\right]\nonumber\\
    &\times \int\mathcal{D}V^{(e)}\,\frac{1}{\det'\nabla^{(e)\,\alpha}\bar{D}_{\alpha}}\,\exp i \int \mathrm{d}^{4}\theta\,\int\,\mathrm{d}^{3}x\,\mathrm{Tr} \left[\frac{i}{2\pi}P\left[\Gamma^{(e)\,\alpha},\,\bar{D}_{\alpha}V'^{(m)}\right]\right].
\end{align}
Generally here, the integral over $\mathcal{D}V^{(e)}$ is difficult. In the Abelian case, $\nabla^{(e)\,\alpha}\to D^{\alpha}$, so the operator $\nabla^{(e)\,\alpha}\bar{D}_{\alpha}$ is independent of $V^{(e)}$. Its reduced determinant is therefore a field-independent constant, and the integrations over $\mathcal{D}V^{(e)}$ and $\mathcal{D}P$ reproduce the Abelian dual photon. In the non-Abelian case, there is an infinite tower of perturbative corrections away from the pure quadratic term $V_{(m)}'^{2}$ and the Abelian flatness constraint $\bar{D}^{\alpha}D_{\alpha}V'^{(m)} = 0$.

\subsubsection{Non-Abelian $\mathcal{N}=2$ duality equations}
To extract the non-Abelian duality equations, we vary the master action with respect to the superfields. The key departure from the Abelian case is that the non-Abelian field strength is a nonlinear functional of its argument; in the master action, this dependence enters through the electric superfield $V^{(e)}$, so its variation requires special care. 

In the path integral, the variation generated by the measure is the linear variation $\delta V$. This variation does not transform covariantly, and therefore obscures gauge covariance when varying the non-Abelian field strength. To keep covariance manifest, we introduce the adjoint-valued covariant fluctuation,
\begin{equation}
    \label{Eq: Covariant variation definition}
    \Delta V:=e^{-V}\,\delta e^{V},
\end{equation}
which transforms as $\Delta V\to g^{\dagger}\Delta V g$. In terms of $\Delta V$, the variation of the connection is
\begin{equation}
    \label{eq:deltaGamma-cov-short}
    \delta\Gamma_\alpha = D_\alpha\Delta V + [\,\Gamma_\alpha,\,\Delta V\,]\equiv \nabla_\alpha\Delta V,\qquad
    \nabla_\alpha:=D_\alpha+[\Gamma_\alpha,\,\cdot].
\end{equation}
Since $W = \frac{i}{2}\bar{D}^{\alpha}\Gamma_{\alpha}\left(V\right)$, the field strength varies as
\begin{equation}
    \delta W(V)
    = \frac{i}{2}\,\bar D^{\alpha}\!\big(\nabla_\alpha \Delta V\big).
\end{equation}
These will be used to derive the non-Abelian duality equations.

To connect the linear variation appearing in the measure with the covariant fluctuation appearing in $\delta W$, we treat the map $\delta V\to \Delta V$ as a linear operator built from nested commutators. In this language, \eqref{Eq: Covariant variation definition} becomes
\begin{align}
    \Delta V^{(e)}&=\frac{1-e^{-\left[V^{(e)},\,\cdot\right]}}{\left[V^{(e)},\,\cdot\right]}\,\delta V^{(e)}.
\end{align}
Expanding the operator as a power series then gives
\begin{align}
    \Delta V^{(e)}&= \sum_{n=0}^{\infty}\frac{(-1)^{n}\left[V^{(e)},\,\cdot\right]^{n}\,\delta V^{(e)}}{\left(n+1\right)!}.
\end{align}
The relation may be inverted using
\begin{equation}
    \delta V^{(e)}=\frac{\left[V^{(e)},\,\cdot\right]}{\,1-e^{-\left[V^{(e)},\,\cdot\right]}\,}\,\Delta V^{(e)},
\end{equation}
whose power series expansion involves the Bernoulli numbers $B_{n}$: 
\begin{align}
    \delta V^{(e)} &= \sum_{n=0}^{\infty}\frac{\left(-1\right)^{n}B_{n}\left[V^{(e)},\,\cdot\right]^{n}\,\Delta V^{(e)}}{n!}.
\end{align}
Finally, we also use 
\begin{equation}
    \mathrm{Tr}\big(X\,[V,\cdot]^n Y\big)
    \;=\;(-1)^n\,\mathrm{Tr}\big([V,\cdot]^n X\,Y\big),
\end{equation}
which follows from the cyclicity of the trace. Using these relations, we may express the linear variation of each term in the master action in a form that factorises against the covariant fluctuation of the underlying fields. 

For a linear variation of $V^{(e)}$, the contribution from the mixing term varies as 
\begin{align}
    &\delta \left(\frac{2i}{\pi}\mathrm{Tr}\int\mathrm{d}^{3}x\,\int\mathrm{d}^{4}\theta \,V'^{(m)}W^{(e)}\right)\nonumber\\& = \frac{i}{\pi}\mathrm{Tr}\int\mathrm{d}^{3}x\,\int\mathrm{d}^{4}\theta \,\left(-i\bar{D}^{\alpha}D_{\alpha}V'^{(m)} + \left[V'^{(m)},\,W^{(e)}\right]-i\bar{D}^{\alpha}\left[V'^{(m)},\,\Gamma_{\alpha}^{(e)}\right]\right)\Delta V^{(e)},
\end{align} 
and the Chern--Simons term varies as 
\begin{align}
    \delta S_{CS}
    &= \frac{ik}{\pi}\mathrm{Tr}\!\int d^{3}x\,d^{4}\theta\,
    \left(\,
     \frac{i}{2}\,\bar D^{\alpha}D_{\alpha}V^{(e)}
     - \frac{i}{2}\,\bar D^{\alpha}\big[\,V^{(e)},\Gamma_{\alpha}^{(e)}\,\big]
     \right.\nonumber\\&\qquad\qquad\qquad\qquad\qquad\left.+ \frac{1}{2}[\,V^{(e)},W^{(e)}\,]
     + \frac{\left[V^{(e)},\,\cdot\right]}{\,e^{\left[V^{(e)},\,\cdot\right]}-1\,}\,W^{(e)}
    \right)\,\Delta V^{(e)}.
\end{align}
Similarly, for a linear variation of $V'^{(m)}$, expressed in terms of the covariant variation $\Delta V'^{(m)}$, the mass term varies as
\begin{equation}
    \delta\left(\frac{2g^{2}}{\left(2\pi\right)^{2}}\int \mathrm{d}^{3}x\,\int\mathrm{d}^{4}\theta\,\mathrm{Tr}\left(V_{(m)}'^{2}\right)\right) = \frac{2g^{2}}{\left(2\pi\right)^{2}}\int \mathrm{d}^{3}x\,\int\mathrm{d}^{4}\theta\,\mathrm{Tr}\left(2V'^{(m)}\Delta V'^{(m)}\right),
\end{equation}
and the mixing term varies as
\begin{equation}
    \delta\left( \frac{2i}{\pi}\int \mathrm{d}^{3}x\,\int\mathrm{d}^{4}\theta\,\mathrm{Tr}\left(V'^{(m)}W^{(e)}\right)\right) = \frac{2i}{\pi}\int \mathrm{d}^{3}x\,\int\mathrm{d}^{4}\theta\,\mathrm{Tr}\left(W^{(e)}\Delta V'^{(m)}\right).
\end{equation}
Given these relations, we now vary the master action. Varying linearly with respect to $V'^{(m)}$, for arbitrary covariant fluctuation $\Delta V'^{(m)}$, gives
\begin{equation}
    g^{2}V'^{(m)} =- 2\pi i\, W^{(e)},
\end{equation}
which is identical in form to the Abelian result. Varying linearly with respect to $V^{(e)}$, again for arbitrary $\Delta V^{(e)}$, we obtain
\begin{align}
    &\left(-i\bar{D}^{\alpha}D_{\alpha}V'^{(m)} + \left[V'^{(m)},\,W^{(e)}\right]-i\bar{D}^{\alpha}\left[V'^{(m)},\,\Gamma_{\alpha}^{(e)}\right]\right)\nonumber\\ +& \frac{k}{2}\left(i\bar{D}^{\alpha}D_{\alpha}V^{(e)} + \left[V^{(e)},\,W^{(e)}\right]-i\bar{D}^{\alpha}\left[V^{(e)},\,\Gamma_{\alpha}^{(e)}\right] +2\frac{\left[V^{(e)},\,\cdot\right]}{e^{\left[V^{(e)},\,\cdot\right]}-1}W^{(e)}\right) = 0.
\end{align}
For $g^{2}k\neq0$, while the second equation is not algebraically invertible due to its nested commutator structure, the pair of dual constraint equations nevertheless determines the electric-magnetic dictionary uniquely: the first expresses $V'^{(m)}$ in terms of $W^{(e)}$, and the second imposes a covariant nonlinear constraint that fixes $W^{(e)}$ from a given $V'^{(m)}$. The resulting map is implicit but complete, and provides the full non-Abelian generalisation of the Abelian duality relations. The first equation retains the simple Abelian algebraic structure, while the second encodes the genuinely non-Abelian commutator corrections. Together, these relations fully specify the operator relations between the electric and magnetic descriptions.

For $g^{2}k\neq 0 $, while the inverse map from magnetic observables back to electric ones requires solving the covariant nonlinear constraint rather than an elementary algebraic inversion, the coupled duality equations still determine the correspondence uniquely at the level of operator expectation values; the duality is therefore fully well defined, albeit only implicitly so away from the Abelian limit. At $g^{2}=0$, the first equation reads $W^{(e)}=0$, and the relations reduce to flatness constraints rather than an invertible map.

\subsection{\texorpdfstring{$\mathcal{N}=1$ non-Abelian case}
                        {N=1 non-Abelian case}}
In the Abelian theory, the $\mathcal{N}=1$ master action was obtained by reducing the Abelian $\mathcal{N}=2$ parent and then working directly in $\mathcal{N}=1$ superspace. The non-Abelian $\mathcal{N}=1$ master action is obtained simply by promoting all $\mathcal{N}=1$ superfields to take values in the Lie algebra and replacing the Abelian field strength by its non-Abelian counterpart. 

Concretely, we take
\begin{equation}
    \Gamma_{\alpha} = \Gamma_{\alpha}^{A}T^{A},
\end{equation}
with $T^{A}$ the group generators. Gauge transformations are implemented via
\begin{equation}
    K = K^{A}T^{A},
\end{equation}
where under a transformation, 
\begin{equation}
    \Gamma_{\alpha}\rightarrow e^{-K}\big(\Gamma_{\alpha}+D_{\alpha}\big)e^{K},
    \qquad
    \delta_{K}\Gamma_{\alpha}=D_{\alpha}K + [\Gamma_{\alpha},K].
\end{equation}
The non-Abelian $\mathcal{N}=1$ field strength is 
\begin{equation}
    W_{\alpha}\left(\Gamma\right)= D^{\beta}D_{\alpha}\Gamma_{\beta} - \left\{\Gamma^{\beta},\,D_{\beta}\Gamma_{\alpha}\right\}  - \frac{1}{3}\left\{\Gamma^{\beta},\,\left\{\Gamma_{\beta},\,\Gamma_{\alpha}\right\}\right\}.
\end{equation}
This obeys the Bianchi identity $\nabla^{\alpha}W_{\alpha}=0$ and transforms covariantly,
\begin{equation}
    \delta_{K}W_{\alpha}=[W_{\alpha},\,K],
    \qquad
    \nabla_{\alpha}X = D_{\alpha}X + [\Gamma_{\alpha},\,X].
\end{equation}
The non-Abelian Stückelberg multiplet is the $\mathcal{N}=1$ descendant of the $\mathcal{N}=2$ compensator
\begin{equation}
    \Gamma'^{\alpha\,(m)} = \Gamma^{(m)}_{\alpha} - \Lambda_{(1)\,\alpha}^{(S)},
\end{equation}
where $\Lambda_{\alpha}$ is induced by $\Sigma_{S} = \Sigma_{S}^{A}T^{A}$ via 
\begin{equation}
    \Lambda_{(1)\,\alpha}^{(S)} = \Sigma^{-1\,(S)}D_{\alpha}\Sigma^{S}.
\end{equation}
This leaves $\Gamma'^{(m)}_{\alpha}$ as the $\mathcal{N}=1$ analogue of $V'^{(m)}$.

With these replacements, the non-Abelian $\mathcal{N}=1$ vector master partition function is the straightforward non-Abelianisation of the Abelian one:
\begin{align}
    \hat{\mathcal{Z}}_{\mathcal{N}=1}^{\text{vector}}
    =\int \mathcal{D}\Gamma^{(m)}\,\mathcal{D}\Gamma^{(e)}\,\mathcal{D}\Lambda_{(1)}^{(S)}\,
    \exp\left\{
    i \int \mathrm{d}^{3}x\,\mathrm{d}^{2}\theta\,
    \mathrm{Tr}\Big[
    -\frac{g^{2}}{(2\pi)^{2}}\Gamma'^{\alpha\,(m)}\Gamma'^{(m)}_{\alpha}\right.\nonumber\\
    - \left.\frac{1}{2\pi}\Gamma'^{\alpha\,(m)}W^{(e)}_{\alpha}
    - \frac{k}{4\pi}\Gamma^{(e)\alpha}W^{(e)}_{\alpha}
    \Big]\right\}.
\end{align}
The effective theories obtained by integrating out the electric or magnetic degrees of freedom are the non-Abelian $\mathcal{N}=1$ analogues of the $\mathcal{N}=2$ theories described earlier. For $g^{2}k\neq 0$, one finds non-Abelian $\mathcal{N}=1$ level-$k$ Yang--Mills--Chern--Simons on the electric side, and a massive Deser--Jackiw vector coupled to a level-$k$ Chern--Simons sector on the magnetic side. For $g^{2}=0$, one lands on a constrained $\mathcal{N}=1$ level-$k$ Chern--Simons path integral on the electric side, and a proposed $\mathcal{N}=1$ non-Abelian Gaiotto--Witten-like representation of the same constrained endpoint on the magnetic side \cite{Gaiotto:2008sa}. For $k=0$ the electric side reduces to non-Abelian $\mathcal{N}=1$ Yang--Mills theory, whereas the magnetic side reduces to the infinite tower of deformations of the $\mathcal{N}=1$ principal chiral model.

At the level of the duality equations, the non-Abelian $\mathcal{N}=1$ master action produces relations that are precisely the $\theta-$ descendants of the $\mathcal{N}=2$ non-Abelian duality equations, with no additional structural features.

%% file: sections/5_Conclusions.tex
\section{Conclusions}
\label{Sec: Conclusions}
In this work we have presented a single off-shell $\mathcal{N}=2$ superspace master partition function in three dimensions, built from two massless vector multiplets and a chiral multiplet that plays the role of a Stückelberg compensator. For $g^{2}k\neq 0$, integrating out the magnetic fields produces the $\mathcal{N}=2$ Maxwell--Chern--Simons theory, while integrating over the electric fields yields the Deser--Jackiw--Chern--Simons theory. At the strict point $g^{2}=0$, the construction gives the constrained Chern--Simons/Gaiotto--Witten-like endpoint, while for $k=0$ it reproduces the Maxwell-scalar duality. The same construction reduces to $\mathcal{N}=1$ superspace by splitting into vector and real-linear (scalar) components, so that the $\mathcal{N}=1$ dual pairs follow from the same framework.

A recurring limitation of approaches in the literature (such as in \cite{Deser:1984kw}) is the reliance on local (classical) equations of motion, which inadvertently erases global (and therefore quantum) information, yielding a semiclassical duality rather than a fully quantum one. Our construction incorporates the global data from the outset. In the Abelian theory, the dual scalar must be compact in order to reproduce the full $\mathrm{U}(1)$ holonomy structure, and this compactness is implemented naturally by the Stückelberg field. The non-Abelian case is more rigid: flat connections cannot, in general, be constructed from a single Lie-algebra-valued multiplet. For this reason, the Stückelberg compensator is taken to be group-valued: it forces the magnetic variables to live in the gauge group $G$ and ensures that, when flatness is imposed, the full space of flat $G$-connections with their correct holonomy sectors appears automatically. This gives a genuinely quantum duality on both sides. In this way, the master construction incorporates the global data of the dual theories before any equations of motion are used. 

For $g^{2}k\neq0$, the upgraded non-Abelian master construction relates Yang--Mills--Chern--Simons theory to a massive adjoint vector superfield coupled to a Chern--Simons term. It reduces correctly to a constrained Chern--Simons path integral or pure Yang--Mills on the electric side at $g^{2}=0$ and $k=0$, respectively. On the magnetic side at $g^{2}=0$, the master partition function gives a proposed non-Abelian Gaiotto--Witten-like representation of the same constrained endpoint. The most difficult regime is the non-Abelian $k=0$ case: while the electric description becomes pure Yang--Mills, the magnetic side turns into an infinite tower of commutator interactions subject to a covariant flatness condition, resembling a highly deformed principal chiral model. We do not obtain a closed-form magnetic dual in this regime; the magnetic description is only implicit, and understanding its precise structure remains an open problem. Beyond the Witten index check, we have not yet analysed dynamical observables, supersymmetric indices, or RG flows between different $\left(g^{2},\,k\right)$ regions, and our construction is restricted to $\mathrm{U}(N)$-type gauge groups without coupling to matter multiplets.

Several natural extensions suggest themselves. The non-Abelian $k=0$ regime is particularly subtle: the magnetic theory may admit a more compact formulation. A systematic study of observables, such as correlation functions, supersymmetric indices, and RG flows between different regions of parameter space, would help clarify how the duality behaves away from the regimes analysed here. From the supersymmetric perspective, it would also be interesting to extend the master construction to three-dimensional $\mathcal{N}=4$. From the gauge-theory perspective, the analysis could be extended to other gauge groups, such as $\mathrm{SO}\left(N\right)$ or $\mathrm{Sp}\left(N\right)$, and coupling to matter multiplets would allow one to track how the duality acts on flavour. Our framework may also clarify aspects of the new dualities recently suggested in \cite{Benvenuti:2025huk}. All of these directions admit brane-configuration realisations in string theory, and pursuing them could enhance our understanding of how the duality is embedded in brane dynamics and how the master construction fits into the broader web of dualities.

%% file: appendices/N=2components_duality.tex
\section{\texorpdfstring{Duality relations for Abelian $\mathcal{N}=2$}
                         {Duality relations for Abelian N=2}}
\label{Appendix: N=2 Schwinger Dyson}
In components,
\begin{equation}
    g^{2}\left(V^{(m)}-\Lambda^{(S)}\right)+2\pi iW^{(e)}= 0
\end{equation}
for insertions independent of $V^{(m)}-\Lambda^{(S)}$ becomes
\begin{align}
     g^{2}C^{(S)} &=- 2\pi \,\phi^{(e)} , \label{eq: map1}\\
     g^{2}\chi^{(S)}  &= 2\pi\,\bar{\lambda}^{(e)} ,\\
     g^{2}\bar{\chi}^{(S)} &=2\pi\,\lambda^{(e)} ,\\
     g^{2}N^{(S)} &=0,\\
     g^{2}N^{\dagger\,(S)} &=0,\\
     g^{2}\left(i\gamma^{m}\left(A_{m}^{(m)}-A_{m}^{(S)}\right) + \varepsilon\phi^{(m)}\right) &=  2\pi\left(\varepsilon D^{(e)} + i\gamma^{r}g_{rl}\varepsilon^{lmn}\partial_{m}A_{n}^{(e)}\right) ,\\
     g^{2}\left(\bar{\lambda}^{(m)}-\frac{i}{2}\slashed{\partial}\chi^{(S)}\right) &=-i\pi  \,\slashed{\partial}\bar{\lambda}^{(e)} ,\\
     g^{2}\left(\lambda^{(m)}+\frac{i}{2}\slashed{\partial}\bar{\chi}^{(S)}\right) &= i\pi\,\slashed{\partial}\lambda^{(e)} ,\\
     g^{2}\left(D^{(m)} - \frac{1}{2}\partial^{2}C^{(S)}\right) &=-\pi\,\partial^{2}\phi^{(e)}   \label{eq: map9}.
\end{align}
Similarly, for insertions independent of both $V^{(m)}-\Lambda^{(S)}$ and $V^{(e)}$, 
\begin{equation}
     B =  V^{(e)} - \frac{i}{M}W^{(e)}
\end{equation}
becomes
\begin{align}
     C^{(B)}  &=  C^{(e)} + \frac{1}{M}\phi^{(e)} ,\\
     \chi^{(B)} &= \chi^{(e)} - \frac{1}{M}\bar{\lambda}^{(e)} ,\\
     \bar{\chi}^{(B)} &= \bar{\chi}^{(e)}+\frac{1}{M}\lambda^{(e)} ,\\
     N^{(B)} &= N^{(e)} ,\\
     N^{\dagger\,(B)} &= N^{\dagger\,(e)} ,\\
     i\gamma^{m}A_{m}^{(B)} + \varepsilon\phi^{(B)} &= i\gamma^{m}A_{m}^{(e)} + \varepsilon\phi^{(e)}+\frac{1}{M}\left(\varepsilon D^{(e)}+i\gamma_{l}\varepsilon^{lmn}\partial_{m}A_{n}^{(e)}\right) ,\\
     \bar{\lambda}^{(B)}+\frac{i}{2}\slashed{\partial}\chi^{(B)} &=\bar{\lambda}^{(e)}+\frac{i}{2}\slashed{\partial}\chi^{(e)} - \frac{i}{2M}\slashed{\partial}\bar{\lambda}^{(e)} ,\\
     \lambda^{(B)} - \frac{i}{2}\slashed{\partial}\bar{\chi}^{(B)} &=\lambda^{(e)} - \frac{i}{2}\slashed{\partial}\bar{\chi}^{(e)} + \frac{i}{2M}\slashed{\partial}\lambda^{(e)} ,\\
     D^{(B)}+\frac{1}{2}\partial^{2}C^{(B)} &= D^{(e)}+\frac{1}{2}\partial^{2}C^{(e)} - \frac{1}{2M}\partial^{2}\phi^{(e)} .\label{eq: map final}
\end{align}
These two sets of equations determine fully the magnetic side correlators from the electric ones in the massive case. We then invert this map using
\begin{equation}
    V^{(e)} = B - \frac{1}{k}\left(V^{(m)}-\Lambda^{(S)}\right)
\end{equation}
to get 
\begin{align}
    C^{(e)} &= C^{(B)} + \frac{1}{k}C^{(S)};\\
    \chi^{(e)} &= \chi^{(B)}+ \frac{1}{k}\chi^{(S)}\\
    \bar{\chi}^{(e)}&=\bar{\chi}^{(B)}+\frac{1}{k}\bar{\chi}^{(S)}\\
    N^{(e)}&=N^{(B)}+ \frac{1}{k}N^{(S)}\\
    N^{\dagger\,(e)}&=N^{\dagger\,(B)}+ \frac{1}{k}N^{\dagger\,(S)}\\
    i\gamma^{m}A_{m}^{(e)} + \varepsilon\phi^{(e)}&=i\gamma^{m}A_{m}^{(B)}+\varepsilon\phi^{(B)}- \frac{1}{k}\left(i\gamma^{m}\left(A_{m}^{(m)}-A_{m}^{(S)}\right)+\varepsilon\phi^{(m)}\right)\\
    \bar{\lambda}^{(e)}+\frac{i}{2}\slashed{\partial}\chi^{(e)}&=\bar{\lambda}^{(B)}+\frac{i}{2}\slashed{\partial}\chi^{(B)}- \frac{1}{k}\left(\bar{\lambda}^{(m)}-\frac{i}{2}\slashed{\partial}\chi^{(S)}\right)\\
    \lambda^{(e)} -\frac{i}{2}\slashed{\partial}\bar{\chi}^{(e)}&=\lambda^{(B)} -\frac{i}{2}\slashed{\partial}\bar{\chi}^{(B)}- \frac{1}{k}\left(\lambda^{(m)} +\frac{i}{2}\slashed{\partial}\bar{\chi}^{(S)}\right)\\
    D^{(e)}+ \frac{1}{2}\partial^{2}C^{(e)}&= D^{(B)}+ \frac{1}{2}\partial^{2}C^{(B)}-\frac{1}{k}\left(D^{(m)}- \frac{1}{2}\partial^{2}C^{(S)}\right) \label{Eq: Final mapping}.
\end{align}
Equations \eqref{eq: map1} to \eqref{Eq: Final mapping} provide the full component--level map between each side of the duality for $g^{2}k\neq 0$.

%% file: appendices/N=1masterfromN=2master.tex
\section{\texorpdfstring{$\mathcal{N}=1$ Master from the $\mathcal{N}=2$ Master}
                         {N=1 Master from the N=2 Master}}
\label{Appendix: N=1 from N=2}
To reduce to $\mathcal{N}=1$ supersymmetry, we introduce the chiral coordinate $y^{m} = x^{m}+i\theta\gamma^{m}\bar{\theta}$, so that at fixed $\left(y,\,\theta\right)$, 
\begin{equation}
    D_{\alpha} = \frac{\partial}{\partial\theta^{\alpha}} + 2i\gamma^{m}_{\alpha\beta}\bar{\theta}^{\beta}\partial_{m}
\end{equation}
and
\begin{equation}
    \bar{D}_{\alpha} = -\frac{\partial}{\partial\bar{\theta}^{\alpha}}.
\end{equation}
Using these, we have that for a generic
\begin{equation}
    F = f + \bar{\theta}^{\alpha}f_{\alpha}+ \bar{\theta}^{2}f_{2},
\end{equation}
hence
\begin{equation}
    \int \mathrm{d}^{2}\bar{\theta}\,F = f_{2}.
\end{equation}
At fixed $y$, we have
\begin{equation}
    \bar{D}^{2}F \;\bigg|_{\bar{\theta}=0}=-4\int \mathrm{d}^{2}\bar{\theta}\,F,
\end{equation}
and so
\begin{equation}
    \int \mathrm{d}^{4}\theta\, F = -\frac{1}{4}\int \mathrm{d}^{2}\theta\,\bar{D}^{2}F\;\bigg|_{\bar{\theta}=0}.
\end{equation}
We make the splitting
\begin{equation}
    V=V_0+\bar\theta^\alpha\Gamma_\alpha+\bar\theta^2 S
\end{equation}
with
\begin{equation}
    V_0=V\Big|_{\bar{\theta}=0},\qquad
    \Gamma_\alpha=\frac{\partial V}{\partial\bar\theta^\alpha}\Big|_{\bar{\theta}=0},\qquad
    S=\frac14\frac{\partial^2V}{\partial\bar\theta^\alpha\partial\bar\theta_\alpha}\Big|_{\bar{\theta}=0}.
\end{equation}
The useful $\mathcal{N}=1$ blocks are
\begin{equation}
    \Sigma=\frac{1}{2} D^\alpha\Gamma_\alpha,\qquad
    W_\alpha=\tfrac12 D^\beta D_\alpha\Gamma_\beta.
\end{equation}
The prepotential shift $\Gamma_\alpha\to\Gamma_\alpha+D_\alpha K$ leaves $W_\alpha$ invariant and shifts $\Sigma\to\Sigma+\frac{1}{2} D^2K$.  Note the relation
\begin{align}
    W^{(e)}\bigg|_{\bar{\theta}=0} = -i\Sigma^{(e)}\bigg|_{\bar{\theta}=0} .
\end{align}
We also have that
\begin{equation}
    D_{\alpha}W\bigg|_{\bar{\theta}=0}=iW_{\alpha}\bigg|_{\bar{\theta}=0}.
\end{equation}
Similarly
\begin{align}
    \bar{D}_{\alpha}W\bigg|_{\bar{\theta}=0} &=\gamma^{m}_{\alpha\beta}\partial_{m}\Gamma^{\beta}\bigg|_{\bar{\theta}=0}+iD_{\alpha}S\bigg|_{\bar{\theta}=0}.
\end{align}
A useful relation is the product rule for bosonic $A$ and $B$
\begin{equation}
    \bar{D}^{2}\left(AB\right)=\left(\bar{D}^{2}A\right)B+2\left(\bar{D}^{\alpha}A\right)\left(\bar{D}_{\alpha}B\right)+A\left(\bar{D}^{2}B\right).
\end{equation}
Similarly, for fermionic $A_{\gamma}$ and $B_{\gamma}$, 
\begin{equation}
    \bar{D}^{2}\left(A^{\gamma}B_{\gamma}\right)=\left(\bar{D}^{2}A^{\gamma}\right)B_{\gamma}-2\left(\bar{D}^{\alpha}A^{\gamma}\right)\left(\bar{D}_{\alpha}B_{\gamma}\right)+A^{\gamma}\left(\bar{D}^{2}B_{\gamma}\right)
\end{equation}
Using that for any function of superspace $F$, we may write
\begin{equation}
    F\left(y,\,\theta,\,\bar{\theta}\right) = F| -\bar{\theta}^{\alpha}\bar{D}_{\alpha}F| + \frac{1}{4}\bar{\theta}^{2}\bar{D}^{2}F.
\end{equation}
We apply this to $W$ in chiral coordinates, this yields
\begin{equation}
    W\left(y,\,\theta,\,\bar{\theta}\right) = W\bigg|_{\bar{\theta}=0} - \bar{\theta}^{\alpha}\bar{D}_{\alpha}W\bigg|_{\bar{\theta}=0}.
\end{equation}
From here we have that
\begin{equation}
    W\bigg|_{\bar{\theta}=0} = -i\Sigma,
\end{equation}
which can be verified by computing components. Similarly, we can find the $\bar{\theta}$ component
\begin{align}
    W = \frac{i}{2}\varepsilon^{\beta\gamma}\bar{D}_{\gamma}\left(D_{\beta}V\right)&= \frac{i}{2}\varepsilon^{\beta\gamma}\bar{D}_{\gamma}\left(D_{\beta}\left(V_{0} + \bar{\theta}^{\alpha}\Gamma_{\alpha}+\bar{\theta}^{2}S\right)\right)\\
    &=\ldots -i\bar{\theta}^{\beta}\left(D_{\beta}S +i\varepsilon^{\delta\alpha}\gamma^{m}_{\beta\delta}\partial_{m}\Gamma_{\alpha}\right),
\end{align}
Putting these together, 
\begin{equation}
    W = -i\Sigma - \bar{\theta}^{\alpha}\left(iD_{\alpha}S-\gamma^{m}_{\alpha\beta}\partial_{m}\Gamma^{\beta}\right).
\end{equation}
We also have that 
\begin{equation}
    \label{Eq: Contraction identity}
    \int \mathrm{d}^{2}\theta\,\Gamma^{(m)\,\alpha}D_{\alpha}S^{(e)}\bigg|_{\bar{\theta}=0} = \int \mathrm{d}^{2}\theta\,\left[i\Gamma^{(m)\,\alpha}\gamma^{m}_{\alpha\beta}\partial_{m}\Gamma^{(e)\,\beta}-\Gamma^{(m)\,\alpha}W^{(e)}_{\alpha}\right]\bigg|_{\bar{\theta}=0},
\end{equation}
With these, we may reduce the three terms in the master action. Throughout, a prime denotes the Stückelberg shifted field $V' = V - \Lambda^{(S)}$, with $V_{0}'$, $\Gamma_{\alpha}'$, and $S'$ the corresponding $\mathcal{N}=1$ components.

Firstly, the Stückelberg mass
\begin{align}
    \int \mathrm{d}^{4}\theta\,V'^{2} &= - \frac{1}{4}\int \mathrm{d}^{2}\theta\,\bar{D}^{2}\left(V'^{2}\right)\bigg|_{\bar{\theta}=0}\\
    &=-\frac{1}{2}\int \mathrm{d}^{2}\theta\,\left(V'\bar{D}^{2}V'+\left(\bar{D}^{\alpha}V'\right)\left(\bar{D}_{\alpha}V'\right)\right)\bigg|_{\bar{\theta}=0}.
\end{align}
Dealing with the two terms separately, we then have
\begin{align}
    -\frac{1}{2}\int \mathrm{d}^{2}\theta\,\left(V'\bar{D}^{2}V'\right)|_{\bar{\theta}=0} &= -\frac{1}{2}\int \mathrm{d}^{2}\theta\,\left(\left(V_{0}' + \bar{\theta}^{\alpha}\Gamma_{\alpha}' + \bar{\theta}^{2} S'\right)\bar{D}^{2}\left(V_{0}' + \bar{\theta}^{\alpha}\Gamma_{\alpha}' + \bar{\theta}^{2} S'\right)\right)|_{\bar{\theta}=0}\\
    &=2\int \mathrm{d}^{2}\theta\,V_{0}'S'.
\end{align}
Then also
\begin{align}
    -\frac{1}{2}\int \mathrm{d}^{2}\theta\,\left(\left(\bar{D}^{\alpha}V'\right)\left(\bar{D}_{\alpha}V'\right)\right)\bigg|_{\bar{\theta}=0} 
    &=-\frac{1}{2}\int \mathrm{d}^{2}\theta\,\left(\left(-\Gamma'^{\alpha}-2\bar{\theta}^{\alpha} S' \right)\left(-\Gamma'_{\alpha}-2\bar{\theta}_{\alpha} S'\right)\right)\bigg|_{\bar{\theta}=0}\\
    &=-\frac{1}{2}\int \mathrm{d}^{2}\theta\,\Gamma'^{\alpha}\Gamma'_{\alpha}.
\end{align}
Putting these together, we have
\begin{equation}
    \int \mathrm{d}^{4}\theta\,V'^{2} = \int \mathrm{d}^{2}\theta\,\left(-\frac{1}{2}\Gamma'^{\alpha}\Gamma'_{\alpha} + 2V_{0}'S'\right).
\end{equation}
Then, the mixing term
\begin{align}
    \int \mathrm{d}^{4}\theta\,V^{(m)}W^{(e)} &=\int \mathrm{d}^{4}\theta\,\left[\left(V_{0}^{(m)}+\bar{\theta}^{\alpha}\Gamma_{\alpha}^{(m)}+\bar{\theta}^{2}S^{(m)}\right)\left(-i\Sigma^{(e)}-\bar{\theta}^{\beta}\left(iD_{\beta}S^{(e)} - \gamma^{m}_{\beta\gamma}\partial_{m}\Gamma^{(e)\,\gamma}\right)\right)\right]\\
    &=\int \mathrm{d}^{2}\theta\,\left[-iS^{(m)}\Sigma^{(e)}+\frac{1}{2}\Gamma^{(m)\,\alpha}\left(iD_{\alpha}S^{(e)}-\gamma^{m}_{\alpha\beta}\partial_{m}\Gamma^{(e)\,\beta}\right)\right],
\end{align}
then using \eqref{Eq: Contraction identity}, we have
\begin{align}
    \int \mathrm{d}^{4}\theta\,V^{(m)}W^{(e)} &=\int \mathrm{d}^{2}\theta\,\left[-iS^{(m)}\Sigma^{(e)} -\Gamma^{(m)\,\alpha}\gamma^{m}_{\alpha\beta}\partial_{m}\Gamma^{(e)\,\beta} + \frac{i}{4}\Gamma^{(m)\,\alpha}W_{\alpha}^{(e)}\right]
\end{align}
By writing
\begin{equation}
    \int \mathrm{d}^{4}\theta\,V^{(m)}W^{(e)} =\frac{1}{2}\int \mathrm{d}^{4}\theta\,\left[V^{(m)}W^{(e)}+V^{(e)}W^{(m)}\right],
\end{equation}
we see that there is a decoupling up to the boundary term of the prepotential.
\begin{align}
    &\int \mathrm{d}^{4}\theta\,V^{(m)}W^{(e)}\\=&\nonumber\int \mathrm{d}^{2}\theta\,\left[-\frac{i}{2}\left(S^{(m)}\Sigma^{(e)}+S^{(e)}\Sigma^{(m)}\right) + \frac{i}{8}\left(\Gamma^{(m)\,\alpha}W_{\alpha}^{(e)} + \Gamma^{(e)\,\alpha}W_{\alpha}^{(m)} \right) - \partial_{m}\left(\Gamma^{(e)}\gamma^{m}\Gamma^{(m)}\right)\right].
\end{align}
This boundary term does not vanish in general, however, we may calculate it and see that it is
\begin{equation}
    -\int \mathrm{d}^{2}\theta\,\partial_{m}\left(\Gamma^{(m)}\gamma^{m}\Gamma^{(e)}\right) = -i\partial_{m}\left(\varepsilon^{nmr}A_{n}^{(m)}A_{r}^{(e)}\right).
\end{equation}
Interestingly, this does not vanish in general, however when $(m)\rightarrow (e)$, it will vanish. We discard this term regardless by working on closed manifolds.

For the Chern--Simons term, the mixing term reduces upon $(m)\rightarrow (e)$ to
\begin{align}
    \int \mathrm{d}^{4}\theta\,V^{(e)}W^{(e)}=&\int \mathrm{d}^{2}\theta\,\left[-iS^{(e)}\Sigma^{(e)} + \frac{i}{4}\Gamma^{(e)\,\alpha}W_{\alpha}^{(e)}\right].
\end{align}
The $\mathcal{N}=2$ master action
\begin{equation}
    S = \int \mathrm{d}^{3}x\,\int\mathrm{d}^{4}\theta\,\left[\frac{g^{2}}{\left(2\pi\right)^{2}}V'^{2}_{(m)}
    +\frac{{2i}}{2\pi}\,V^{'(m)}W^{(e)}
    +\frac{ik}{2\pi}V^{(e)}W^{(e)}\right]
\end{equation}
then reduces to its $\mathcal{N}=1$ counterpart
\begin{align}
    S &= \int \mathrm{d}^{3}x\,\int \mathrm{d}^{2}\theta\,\left[\frac{g^{2}}{\left(2\pi\right)^{2}}\left(-\frac{1}{2}\Gamma'^{(m)\,\alpha}\Gamma^{'(m)}_{\alpha} + 2V_{0}^{'(m)}S^{'(m)}\right)\right]\nonumber\\
    &\qquad+\frac{2i}{2\pi}\left(-iS^{(m)}\Sigma^{(e)} -\Gamma^{(m)\,\alpha}\gamma^{m}_{\alpha\beta}\partial_{m}\Gamma^{(e)\,\beta} + \frac{i}{4}\Gamma^{(m)\,\alpha}W_{\alpha}^{(e)}\right)\nonumber\\
    &\qquad +\frac{ik}{2\pi}\left(-iS^{(e)}\Sigma^{(e)} + \frac{i}{4}\Gamma^{(e)\,\alpha}W_{\alpha}^{(e)}\right).
\end{align}
This then splits into the vector and scalar parts
\begin{align}
    S&=\int \mathrm{d}^{3}x\,\int \mathrm{d}^{2}\theta\,\left[-\frac{g^{2}}{2\left(2\pi\right)^{2}}\Gamma'^{(m)\,\alpha}\Gamma^{'(m)}_{\alpha} - \frac{1}{4\pi}\Gamma'^{(m)\,\alpha}W_{\alpha}^{(e)} - \frac{k}{8\pi}\Gamma^{(e)\,\alpha}W_{\alpha}^{(e)}\right]\nonumber\\
    &+\int \mathrm{d}^{3}x\,\int \mathrm{d}^{2}\theta\,\left[\frac{2g^{2}}{\left(2\pi\right)^{2}}V_{0}^{'(m)}S^{'(m)} + \frac{1}{\pi}S^{'(m)}\Sigma^{(e)} + \frac{k}{2\pi}S^{(e)}\Sigma^{(e)}\right]
\end{align}
as the decoupled $\mathcal{N}=1$ vector and $\mathcal{N}=1$ scalar master actions.

%% file: appendices/nonabelian_interactions.tex
\section{Non-Abelian Field Strength Interaction Expansion}
\label{Appendix:Integrating-the-field-strength-interactions}
We start from the interaction piece that appears when splitting the non-Abelian field strength,
\begin{equation}
  I(X,Y)= -\,\frac{i}{2}\,\bar D^\alpha\!\int_0^1\!ds\int_0^1\!s\,d\tau\Big[
  e^{-s(1-\tau)\mathrm{ad}_X}\,\mathrm{ad}_Y\,e^{-s\tau\,\mathrm{ad}_{X+Y}}\,D_\alpha X
  +(X\leftrightarrow Y)
  \Big],
\end{equation}
where we have the shorthand $\mathrm{ad}_{X}\left(\cdot\right) = \left[X,\,\cdot\,\right]$. Expanding the exponentials and collecting powers of $s$ and $\tau$ gives
\begin{equation}
    e^{\,-s(1-\tau)\operatorname{ad}_{X}}\,\operatorname{ad}_{Y}\,e^{\,-s\tau\operatorname{ad}_{X+Y}}
    \;=\;
    \sum_{n,m\ge 0}\frac{(-1)^{n+m}{s^{n+m}(1-\tau)^{n}\tau^{m}}}{n!\,m!}\;
    \operatorname{ad}_{X}^{\,n}\,\operatorname{ad}_{Y}\,\operatorname{ad}_{X+Y}^{\,m},
\end{equation}
and similarly for $X\leftrightarrow Y$ in the second term.
The $s$- and $\tau$-integrals factor and evaluate to
\begin{equation}
    \int_{0}^{1}\mathrm{d}s\;s^{n+m+1}=\frac{1}{n+m+2},
    \qquad
    \int_{0}^{1}\mathrm{d}\tau\;(1-\tau)^{n}\tau^{m}=\frac{n!\,m!}{(n+m+1)!}.
\end{equation}
This leaves that
\begin{equation}
    \int_{0}^{1} \mathrm{d}s\int_{0}^{1} \mathrm{d}\tau \,\frac{(-1)^{n+m}{s^{n+m+1}(1-\tau)^{n}\tau^{m}}}{n!\,m!}=  \frac{\left(-1\right)^{n+m}}{\left(n+m+2\right)!}.
\end{equation}
The resulting double series is then
\begin{equation}
    \label{Eq:Unfactorised-interaction-term}
    I\left(X,\,Y\right) = -\frac{i}{2}\bar{D}^{\alpha}\sum_{n,\,m=0}^{\infty}\frac{\left(-1\right)^{n+m}}{\left(n+m+2\right)!}\left[\operatorname{ad}_{X}^{n}\operatorname{ad}_{Y}\operatorname{ad}_{X+Y}^{m}\left[D_{\alpha}X\right] + \operatorname{ad}_{Y}^{n}\operatorname{ad}_{X}\operatorname{ad}_{X+Y}^{m}\left[D_{\alpha}Y\right]\right].
\end{equation}
For each fixed $n$, we define
\begin{equation}
 F_{n}(A)\;\equiv\;\sum_{m=0}^{\infty}\frac{(-1)^{m+n}A^{m}}{(m+n+2)!}
\;=\;A^{-(n+2)}\left(e^{-A}-\sum_{k=0}^{n+1}\frac{(-1)^{k}A^{k}}{k!}\right).
\label{eq:F-def}
\end{equation}
Here $A^{-(n+2)}$ is understood via its power-series action on adjoint-valued arguments. Then \eqref{Eq:Unfactorised-interaction-term} becomes
\begin{equation}
I(X,Y)
= -\frac{i}{2}\bar{D}^{\alpha}\sum_{n=0}^{\infty}\Big(
\operatorname{ad}_{X}^{\,n}\operatorname{ad}_{Y}\, F_{n}(\operatorname{ad}_{X+Y})\,[D_{\alpha} X]
\;+\;
\operatorname{ad}_{Y}^{\,n}\operatorname{ad}_{X}\, F_{n}(\operatorname{ad}_{X+Y})\,[ D_{\alpha} Y]
\Big).
\label{eq:I-XY-resummed}
\end{equation}
Defining
\begin{equation}
    K_{n}\left(X,\,Y\right) = \operatorname{ad}_{X}^{n}\operatorname{ad}_{Y},
\end{equation}
we then have that
\begin{equation}
    I(X,Y)
    = -\frac{i}{2}\bar{D}^{\alpha}\sum_{n=0}^{\infty}\Big(
    K_{n}\left(X,\,Y\right)\, F_{n}(\operatorname{ad}_{X+Y})\,[ D_{\alpha} X]
    \;+\;
    K_{n}\left(Y,\,X\right)\, F_{n}(\operatorname{ad}_{X+Y})\,[D_{\alpha} Y]
    \Big).
\end{equation}